\documentclass[paper]{JHEP3}
\usepackage{graphicx}
\usepackage{latexsym}

\usepackage{epsfig,multicol,bbm}

\newcommand\fverb{\setbox\pippobox=\hbox\bgroup\verb}
\newcommand\fverbdo{\egroup\medskip\noindent%
            \fbox{\unhbox\pippobox}\ }
\newcommand\fverbit{\egroup\item[\fbox{\unhbox\pippobox}]}
\newbox\pippobox

\title{Hypermultiplet dependence of one-loop effective action
in the ${\cal N}=2$ superconformal theories}

\author{I.L. Buchbinder\\
    Tomsk State Pedagogical University, Tomsk 634041, Russia\\
    E-mail: \email{joseph@tspu.edu.ru}}
\author{ N.G. Pletnev\\
Department of Theoretical Physics, Institute of Mathematics,
Novosibirsk, 630090, Russia\\
E-mail: \email{pletnev@math.nsc.ru}}

\received{February 20, 2001}        
\revised{May 1, 2001}
\accepted{November 27, 2001}        

\preprint{\hepth{0611145}}  

\abstract{We study the one-loop low-energy effective action in the
hypermultiplet sector for ${\cal N}=2$ superconformal models. Any
such a model contains an ${\cal N}=2$ vector multiplet and some
number of hypermultiplets. Gauge group $G$ is assumed to be broken
down to $\tilde{G}\times K$ where $K$ is an Abelian subgroup and a
background vector multiplet belongs to the Cartan subalgebra
corresponding to $K$. We found a general expression for the
low-energy effective action in the form of a proper-time integral.
The leading space-time dependent contributions to the effective
action are derived and their bosonic component structure is
analyzed. The component action contains terms with three and four
space-time derivatives of component fields and has the
Chern-Simons-like form. }

\keywords{Extended Supersymmetry, Harmonic Superspace, Quantum Gauge
Theories, Effective Action }

\begin{document}

\newcommand{\be}{\begin{equation}}
\newcommand{\ee}{\end{equation}}
\newcommand{\bea}{\begin{eqnarray}}
\newcommand{\eea}{\end{eqnarray}}

\section{Introduction}
Four-dimensional ${\cal N}=2$ supersymmetric gauge theories are
formulated in terms of ${\cal N}=2$ vector multiplet coupled to a
massless hypermultiplets in certain representations ${ R}$ of the
gauge group $G$. All such models possess only one-loop divergences
\cite{hsw}, \cite{w}, \cite{hst}, \cite{bko} and can be made finite
at certain restrictions on representations and field contents. In
the model with $n_\sigma$ hypermultiplets in representations
$R_\sigma$ of the gauge group ${G}$ the finiteness condition has
simple and universal form \cite{hsw}
\begin{equation}\label{fin}
C({G})=\sum_\sigma n_\sigma T({R_\sigma}),
\end{equation}
where $C(G)$ is the quadratic Casimir operator for the adjoint
representation and $T(R_\sigma)$ is the quadratic Casimir operator
for the representation ${R_\sigma}$. A simplest solution to
Eq.(\ref{fin}) is ${\cal N}=4$ SYM theory where $n_\sigma=1$ and all
fields are taken in the adjoint representation. It is evident that
there are other solutions, e.g. for the case of $\mathop{\rm SU}(N)$
group and hypermultiplets in the fundamental representation one gets
$T(R)=1/2$, $C(G) =N$ and $n_\sigma = 2N$. A number of ${\cal N}=2$
superconformal models has been constructed in the context of AdS/CFT
correspondence (see e.g \cite{Ahar}, \cite{kach}, \cite{doug} and
references therein, the examples of such models and description of
structure of vacuum states were discussed in details e.g. in Ref.
\cite{9}). All these theories arise in ${\cal N}=4$ SYM theories on
"orbifolds" of $AdS_5 \times S^5$ with respect to discrete subgroup
$\mathop{\rm SU}(4)$ of $\mathop{\rm R}$-symmetry which breaks some
number of the supersymmetries of the ${\cal N}=4$ SYM theory and
acts nontrivially on the gauge group.

In this paper we study the structure of the low-energy one-loop
effective action for the ${\cal N}=2$ superconformal theories. The
effective action of the ${\cal N}=4$ SYM theory and ${\cal N}=2$
superconformal models in the sector of ${\cal N}=2$ vector multiplet
has been studied by various methods \cite{n2}, \cite{n4}, \cite{9},
\cite{10}, \cite{bkt}, \cite{skim}, \cite{km}. However a problem of
hypermultiplet dependence of the effective action in the above
theories was open for a long time.

The low-energy effective action containing both ${\cal N}=2$ vector
multiplet and hypermultiplet background fields in ${\cal N}=4$ SYM
theory was first constructed in Ref. \cite{bi} and studied in more
details in Refs. \cite{hyper}, \cite{bp}. In this paper we will
consider the hypermultiplet dependence of the effective action for
${\cal N}=2$ superconformal models. Such models are finite theories
as well as the ${\cal N}=4$ SYM theory and one can expect that
hypermultiplet dependence of the effective action in ${\cal N}=2$
superconformal models is analogous to one in ${\cal N}=4$ SYM
theory. However this is not so evident. The ${\cal N}=4$ SYM theory
is a special case of the ${\cal N}=2$ superconformal models, however
it possesses extra ${\cal N}=2$ supersymmetry in comparison with
generic ${\cal N}=2$ models. As it was noted in \cite{bi}, just this
extra ${\cal N}=2$ supersymmetry is the key point for finding an
explicit hypermultiplet dependence of the effective action in ${\cal
N}=4$ SYM theory. Therefore a derivation of the effective action for
${\cal N}=2$ superconformal models in the hypermultiplet sector is
an independent problem.

In this paper we derive the complete ${\cal N}=2$ supersymmetric
one-loop effective action depending both on the background vector
multiplet and hypermultiplet fields in a mixed phase where both
vector multiplet and hypermultiplet have non-vanishing expectation
values\footnote{See example of the effective action on Higgs branch
for ${\cal N}=2$ model in Refs. \cite{ikz}, \cite{gural}}. The
${\cal N}=2$ supersymmetric models under consideration are
formulated in  harmonic superspace \cite{6}, \cite{7}. We develop a
systematic method of constructing the lower- and higher-derivative
terms in the one-loop effective action given in terms of a heat
kernel for certain differential operators on the harmonic superspace
and calculate the heat kernel depending on ${\cal N}=2$ vector
multiplet and hypermultiplet background superfields. We study a
component form of a leading quantum corrections for on-shell and
beyond on-shell background hypermultiplets and find that they
contain, among the others, the terms corresponding to the
Chern-Simons-type actions. The necessity of such manifest scale
invariant $P$-odd terms in effective action of ${\cal N}=4$ SYM
theory, involving both scalars and vectors, has been pointed out in
\cite{tseyt}. Proposal for the higher-derivative terms in the
effective action of the ${\cal N}=2$ models in the harmonic
superspace has been given in \cite{arg}. We show how the terms in
the effective action assumed in \cite{arg} can be actually computed
in supersymmetric quantum field theory.

The paper is organized as follows. Section 2 is devoted to a brief
formulation of the ${\cal N}=2$ supersymmetric models in the
harmonic superspace and description of vacuum structure which we
assume. Also in this section the elements of ${\cal N}=2$
supersymmetric background field method are introduced. In section 3
we discuss the structure of the superspace differential operator
associated with the hypermultiplet dependent one-loop effective
action on a given vacuum. Section 4 is devoted to a straightforward
calculation of the one-loop low-energy effective action for the
background fields satisfying the on-shell conditions (see Egs.
(\ref{onsh})). In addition, the bosonic component effective action
containing terms with four space-time derivatives of scalar
components of the hypermultiplet is derived. The analogous
Chern-Simons-type terms have been discussed in \cite{arg}. In
section 5 we study the possible contributions to the effective
action for the background hypermultiplet, which do not satisfy the
on-shell condition (\ref{onsh}). We show that in the purely bosonic
sector the corresponding contribution contains the Chern-Simons-like
terms with three space-time derivatives analogous to terms proposed
in \cite{arg}. The results are summarized in section 6.

\section{The model and background field splitting}

The harmonic superspace approach provides manifestly covariant
off-shell description of ${\cal N}=2$ supersymmetric field
theories at classical and quantum levels (see modern state of
harmonic superspace approach in \cite{7}). ${\cal N}=2$ harmonic
superspace has been introduced in \cite{6}  extending the standard
${\cal N}=2$ superspace with coordinates
$z^M=(x^m,\theta^\alpha_i, \bar\theta^i_{\dot\alpha})$ ($i =1,2$)
by the harmonics $u^{\pm}_{i}$ parameterizing the two-dimensional
sphere $S^2$: $ u^{+i}u^-_i=1, \quad \overline{u^{+i}}=u^-_i.$

The main advantage of harmonic superspace is that the ${\cal N}=2$
vector multiplet and hypermultiplet can be described by
unconstrained superfields over the analytic subspace with the
coordinates $\zeta^M \equiv (x^m_A,
\theta^{+\alpha},\bar\theta^+_{\dot\alpha}, u^{\pm}_i),$ where the
so-called analytic basis is defined by
\begin{equation}\label{analyt}
x^m_A=x^m-i\theta^+\sigma\bar\theta^-
-i\theta^-\sigma^m\bar\theta^+, \quad
\theta^{\pm}_\alpha=u^{\pm}_i\theta^i_\alpha,\quad
\bar\theta^{\pm}_{\dot\alpha}=u^{\pm}_i\bar\theta^i_{\dot\alpha}~.
\end{equation}
The ${\cal N}=2$ vector multiplet is described by a real  analytic
superfield $V^{++}=V^{++I}(\zeta)T_I$ taking values in the Lie
algebra of the gauge group. A hypermultiplet \cite{sohn},
transforming in the representation $R$ of the gauge group, is
described by an analytic superfield $q^+(\zeta)$ and its conjugate
$\tilde{q}^+(\zeta)$ (see the definition of conjugation in
\cite{7}). The scalar component fields $f^i(x_A)$ of the
hypermultiplet and their conjugate $\bar{f}^i=(f_i)^\dag$ form the
$\mathop{\rm SU}(2)$ doublet. They, as well as the spinor component
fields $\psi_\alpha, \bar\kappa^{\dot\alpha}$ of the hypermultiplet,
appear as the lowest-order components in the $\theta^+,
\bar\theta^+, u^{\pm}_i$ expansion of $q^+, \tilde{q}^+.$ The ${\cal
N}=2$ vector potential superfield $V^{++}$ satisfies the reality
condition $\widetilde{V^{++}}=V^{++}$ with respect to the
generalized conjugation which is the combination of complex
conjugation and the antipodal map and transforms under the gauge
transformations as $\delta V^{++}=-\cal{D}^{++}\lambda$, where
$\lambda$ is an arbitrary real analytic superfield parameter. In the
Wess-Zumino gauge, the superfield $V^{++}$ has a finite number of
the component fields $\phi, \bar\phi, A_m, \lambda_\alpha,
\bar\lambda_{\dot\alpha}, D^{(ij)}$ corresponding to field contents
of  ${\cal N}=2$ vector multiplet \cite{grimm}.

The classical action of ${\cal N}=2$ SYM theory coupled to
hypermultiplets consist of two parts: the pure ${\cal N}=2$ SYM
action  and the $q$-hypermultiplet action in the fundamental or
adjoint representation of the gauge group. Written in the harmonic
superspace \cite{6,7} its action  reads
\begin{equation}\label{class}
S=\frac{1}{2g^2}\mbox{tr}\int d^8z \,{\cal W}^2 +\frac{1}{2}\int d
\zeta^{(-4)} q^{+f}_{a}(D^{++} +igV^{++})q^{+a}_{f}~,
\end{equation}
where we used the doublet notation $q^+_a=(q^+,
-\tilde{q}^+)$\footnote{Hypermultiplet part of the action written in
the symplectic covariant form \cite{7} for any number $n$ of
hypermultiplets $q^+_a=(q^+, -\tilde{q}^+)$, ($a= 1, ..., 2n$). Then
$q^+_a$ is related to $q^{+a}$ by the reality condition
$\widetilde{q^+_a}\equiv q^{+a}=\Omega^{ab}q^+_b$, where
$\Omega^{ab}=\Omega^{ba}$ is the invariant tensor of the symplectic
group, $Sp(N_c)$ for fundamental $q^+$ or $Sp(N^2_c)$ for adjoint
$q^+$ (generalization for more complicated representations looks
evident) and the covariant derivative is denoted as ${\cal
D}^{++}q^{+a}=D^{++}q^{+a} +i {\bf V}^{++a}_{\;\;b} q^{+b}$, ${\bf
V}^{++a}_{\;\;b}=V^{++I}({\bf T}_I)^{a}_{\;\;b}$, ${\bf
T}_I=\pmatrix{-T^T_I&0\cr 0&T_I}.$ We also introduce the flavors as
$N_f$-dimensional vector field $q^{+f}$ so that $q^{+f}_a$ is $N_c
\times N_f$ matrix.}. By construction, the action (\ref{class}) is
manifestly ${\cal N}=2$ supersymmetric. Here $d\zeta^{(-4)}$ denotes
the analytic subspace integration measure and $${\cal
D}^{++}=D^{++}+iV^{++}, \quad
D^{++}=\partial^{++}-2i\theta^+\sigma^m\bar\theta^+\partial_m, \quad
\partial^{++}\equiv u^{+i}\frac{\partial}{\partial u^{-i}}$$ is the
analyticity-preserving covariant harmonic derivative. It can be
shown that $V^{++}$ is the single unconstrained analytic,
$D^+_{(\alpha,\dot\alpha)}V^{++}=0$, prepotential of the pure ${\cal
N}=2$ SYM theory \cite{6}, \cite{7}, and all other geometrical
object are determined in terms of it. So,the covariantly chiral
superfield strength ${\cal W}$ is expressed through the
(nonanalytic) real superfield $V^{--}$ satisfying the equation
$$D^{++}V^{--}-D^{--}V^{++}+i[V^{++}, V^{--}]=0.$$ This equation
has a solution in form of the power series in $V^{++}$ \cite{6},
\cite{ 7}:
$$
V^{--}(z,u)=\sum_{n=1}^\infty \int du_1 ...
du_n(-i)^{n+1}\frac{V^{++}(z,u_1) ...
V^{++}(z,u_n)}{(u^+u^+_1)(u^+_1u^+_2) ... (u_n^+u^+)}~,
$$
and uses  the  harmonic distributions $\frac{1}{u^+_1u^+_2}$ or,
in other words, Green's function\\ $G^{(-1,-1)}(u_1,u_2)$ that
obey the equation
$\partial_1^{++}G^{(-1,-1)}(u_1,u_2)=\delta^{(1,-1)}(u_1,u_2)$.
The rules of integration over $SU(2)$ as well as the properties of
harmonic distributions are given in refs. \cite{6}, \cite{7}. To
simplify the notation, we set $g=1$ at the intermediate stages of
the calculation. The explicit dependence on the coupling constant
will be restored in the final expression for the effective action.

Now the superfield strength is defined by
\begin{equation}\label{str}
{\cal W}=-\frac{1}{4}(\bar{D}^+)^2 V^{--}, \quad \bar{\cal
W}=-\frac{1}{4}(D^+)^2 V^{--}.
\end{equation}
One can prove that the ${\cal W}, \bar{\cal W}$ are gauge invariant,
$u$-independent, ${\cal D}^{\pm\pm}{\cal W}=0$, covariant chiral
(antichiral) superfields, $\bar{\cal D}^{\pm}_{\dot\alpha}{\cal
W}=0$, and satisfy the Bianchi identities $({\cal D}^{\pm})^2{\cal
W}=(\bar{\cal D}^{\pm})^2\bar{\cal W}$. For further use we will
write down also the superalgebra of gauge covariant derivatives with
the notation ${\cal D}^{\pm}_{(\alpha,\dot\alpha)}={\cal
D}^{i}_{(\alpha,\dot\alpha)}u^{\pm}_i$:
\begin{equation}\label{alg}
\{{\cal D}^+_\alpha, {\cal
D}^-_\beta\}=-2i\varepsilon_{\alpha\beta}\bar{\cal W}~, \quad
\{\bar{\cal D}^+_{\dot\alpha}, \bar{\cal
D}^-_{\dot\beta}\}=2i\varepsilon_{\dot\alpha\dot\beta}{\cal W}~,
\end{equation}
$$
\{\bar{\cal D}^+_{\dot\alpha}, {\cal D}^-_\alpha\}=-\{{\cal
D}^+_{\alpha}, \bar{\cal D}^-_{\dot\alpha}\}=2i{\cal
D}_{\alpha\dot\alpha}~,
$$
$$
[{\cal D}^{\pm}_\alpha, {\cal
D}_{\beta\dot\beta}]=\varepsilon_{\alpha\beta}\bar{\cal
D}^{\pm}_{\dot\beta}\bar{\cal W}~, \quad [\bar{\cal
D}^{\pm}_{\dot\alpha}, {\cal
D}_{\beta\dot\beta}]=\varepsilon_{\dot\alpha\dot\beta}{\cal
D}^{\pm}_{\beta}{\cal W}~,
$$
$$
[{\cal D}_{\alpha\dot\alpha}, {\cal
D}_{\beta\dot\beta}]=\frac{1}{2i}\{\varepsilon_{\alpha\beta}\bar{\cal
D}^+_{\dot\alpha}\bar{\cal D}^-_{\dot\beta}\bar{\cal W} +
\varepsilon_{\dot\alpha\dot\beta}{\cal D}^-_{\alpha}{\cal
D}^+_{\beta}{\cal
W}\}=\frac{1}{2i}\{\varepsilon_{\alpha\beta}\bar{F}_{\dot\alpha\dot\beta}
+ \varepsilon_{\dot\alpha\dot\beta}F_{\alpha\beta}\}~.
$$
The operators ${\cal D}^+_\alpha$ and $\bar{\cal
D}^+_{\dot\alpha}$ strictly anticommute
\begin{equation}\label{d+}
\{{\cal D}^+_\alpha,{\cal D}^+_\beta\}=\{\bar{\cal
D}^+_{\dot\alpha},\bar{\cal D}^+_{\dot\beta}\}=\{{\cal
D}^+_\alpha,\bar{\cal D}^+_{\dot\alpha}\}=0~.
\end{equation}
It follows from (\ref{d+}) that ${\cal
D}^+_{(\alpha,\dot\alpha)}=e^{-ib}D^+_{(\alpha,\dot\alpha)}e^{ib}$
for some Lie-algebra-valued real superfield $b(z,u)$ known as the
bridge \cite{6}, \cite{7}. In the so-called $\lambda$-frame
defined by $\Phi^{(p)} \rightarrow e^{ib}\Phi^{(p)}$ the gauge
covariant derivatives ${\cal D}^+_{(\alpha,\dot\alpha)}$ coincide
with the flat derivatives $D^+_{(\alpha,\dot\alpha)}$ and
therefore any covariant analytic superfield becomes a function
over the analytic subspace. A full set of gauge covariant
derivatives includes also the harmonic derivatives $({\cal
D}^{++}, {\cal D}^{--}, {\cal D}^{0})$, which form the algebra
$su(2)$ and satisfy the obviously commutation relations with
${\cal D}^{\pm}_\alpha$ and $\bar{\cal D}^{\pm}_{\dot\alpha}$.
Unlike ${\cal D}^+_{(\alpha,\dot\alpha)}$ the gauge covariant
derivatives ${\cal D}^{\pm\pm}$ acquire connections $V^{\pm\pm}$.

The action (\ref{class}) possesses the superconformal symmetry
$SU(2,2|2)$ which is manifest in the harmonic superspace approach
\cite{7}. The low energy effective action at a generic vacuum of
${\cal N}=2$ gauge theory includes only massless $\mathop{\rm
U}(1)$ vector multiplets and massless neutral hypermultiplets,
since charged vectors and charged hypermultiplets get masses by
the Higgs mechanism. The moduli space of vacua for the theory
under consideration is specified by the following conditions
\cite{8}
\begin{equation}\label{vacua}
[\bar\phi, \phi]=0, \quad \phi f_i=0, \quad \bar{f}^i \bar\phi =0
\quad \bar{f}^{(i} T_I f^{j)}=0~.
\end{equation}
Here the $\phi, \bar\phi$ are the scalar components of ${\cal
N}=2$ vector multiplet and complex scalars $f_{i}$  are the scalar
components of the hypermultiplet.

The structure of a vacuum state is characterized by solutions to
Eqs. (\ref{vacua}). These solutions can be classified according to
the phases or branches of the gauge theory under consideration
\cite{8}. In the pure Coulomb phase $f_i =0$, $\phi \not= 0$ and
unbroken gauge group is $\mathop{\rm U}(1)^{ \rm rank({ G})}$. In
the pure Higgs phase $f_i \not=0$ and the gauge symmetry is
completely broken; there are no massless gauge bosons. It is well
known that F- and D-flatness conditions describing the Higgs branch
can be mapped \cite{mir} to the ADHM constraints determining the
moduli space of instantons. In the mixed phases, i.e. on the direct
product of the Coulomb and Higgs branches (some number of $\phi,
\bar\phi$ is not equal to zero and some number of $f_{i}$ is not
equal to zero) the gauge group is broken down to ${\tilde{G}} \times
K$ where $K$ is some Abelian subgroup and $\mbox{rank}(\tilde{G})$
is reduced in comparison with $\mbox{rank}(G)$.

Further we follow \cite{9} and impose the special restrictions on
the background ${\cal N}=2$ vector multiplet and hypermultiplet.
They are chosen to be aligned along a fixed direction in the moduli
space vacua; in particular, their scalar fields should solve Eqs.
(\ref{vacua}):
\begin{equation}\label{vac}
V^{++}={\bf V}^{++}(\zeta){H}, \quad q^+={\bf q}^+ (\zeta)\Upsilon~.
\end{equation}
Here ${H}$ is a fixed generator in the Cartan subalgebra
corresponding to Abelian subgroup $K$, and $\Upsilon$ is a fixed
vector in the ${ R}$-representation space of the gauge group, where
the hypermultiplet takes values, chosen so that ${H} \Upsilon=0$ and
$\bar\Upsilon {\bf T}_I \Upsilon =0.$ At this point we use notations
from \cite{9}. Eq.(\ref{vac}) defines a single $\mathop{\rm U}(1)$
vector multiplet and a single hypermultiplet which is neutral with
respect to the $\mathop{\rm U}(1)$ gauge subgroup generated by ${
H}$. The freedom in the choice of ${H}$ and $\Upsilon$ can be
reduced by requiring that the field configuration (\ref{vac}) to be
invariant under the maximal unbroken gauge subgroup.

At the tree level and energies below the symmetry breaking scale, we
have free field massless dynamics of the ${\cal N}=2$ vector
multiplet and the hypermultiplet aligned in a particular direction
in the moduli space of vacua. Thus the low energy propagating fields
are massless neutral hypermultiplets and $\mathop{\rm U}(1)$ vector
which form the on shell superfields possessing the properties
\begin{equation}\label{onsh}
(D^{\pm})^2{\cal W}=(\bar{D}^{\pm})^2\bar{\cal W}=0~,
\end{equation}
$$
D^{++}q^{+a}=(D^{--})^2q^{+a}=D^{--}q^{-a}=0, \quad
q^{-a}=D^{--}q^{+a}, \quad D^-_{(\alpha,\dot\alpha)}q^{-a}=0~.
$$
All the notations are given in \cite{7}. The equations
(\ref{onsh}) eliminate the auxiliary fields and put the physical
fields on shell.

At the quantum level, however, exchanges of virtual massive
particles produce the corrections to the action of the massless
fields. The manifestly  ${\cal N}=2$ supersymmetric Feynman rules in
harmonic superspace have been developed in \cite{6} (see also
\cite{7, 7a}). We quantize the ${\cal N}=2$ supergauge theory in the
framework of the ${\cal N}=2$ supersymmetric background field method
\cite{10}, \cite{bko} by splitting the fields $V^{++}, q^{+a}$ into
the sum of the background fields $V^{++}, q^{+a}$, parameterized
according to (\ref{vac}), and the quantum fields $v^{++}, Q^{+a}$
and expanding the Lagrangian in a power series in quantum fields.
Such a procedure allows us to find the effective action for
arbitrary ${\cal N}=2$ supersymmetric gauge model in a form
preserving the manifest ${\cal N}=2$ supersymmetry and classical
gauge invariance in quantum theory. The original infinitesimal gauge
transformations are realized by two different ways: first as the
background transformations:
\begin{equation}\label{cltr}
\delta V^{++}=-{\cal D}^{++}\lambda, \quad \delta
v^{++}=i[\lambda, v^{++}]~,
\end{equation}
and second as the
quantum transformations
\begin{equation}\label{qutr}
\delta V^{++}=0,\quad  \delta v^{++}=-{\cal D}^{++}\lambda
-i[v^{++}, \lambda].
\end{equation}
In the background-quantum splitting, the classical action of the
pure ${\cal N}=2$ SYM theory can be shown to be given by
\begin{equation}\label{splitt}
S_{SYM}[V^{++}+v^{++}]=S_{SYM}[V^{++}]+\frac{1}{4}\int
d\zeta^{(-4)}duv^{++}(D^+)^2{\cal W}_\lambda
\end{equation}
$$
-\mbox{tr}\int d^{12}z \sum_{n=2}^\infty \frac{(-ig)^{n-2}}{n}\int
du_1...du_n\frac{v^{++}_\tau(z,u_1)...v^{++}_\tau(z,u_n)}{(u_1^+u_2^+)...(u_n^+u_1^+)}~.
$$
${\cal W}_\lambda$ and $v^{++}_\tau$ denote the $\lambda$- and
$\tau$-frame forms of $\cal W$ and $v^{++}$ respectively ${\cal
W}_\lambda=e^{ib}{\cal W}e^{-ib}$,
$v^{++}_\tau=e^{-ib}v^{++}e^{ib}$. The quantum part of the action
depends on $V^{++}$ via the dependence of $v^{++}_\tau$ on $b$,
latter being a complicated function of $V^{++}$.  The
hypermultiplet action becomes
\begin{equation}\label{hypeq}
S_H(q+Q)=S_H[q] +\int d\zeta^{(-4)}du Q^+_a{\cal
D}^{++}q^{+a}+\frac{1}{2}\int d\zeta^{(-4)}duq^+_a iv^{++}q^{+a}
\end{equation}
$$
+\frac{1}{2}\int d\zeta^{(-4)}du\{Q^+_a{\cal D}^{++}Q^{+a}+Q^+_a
iv^{++}q^{+a}+q^+_a iv^{++}Q^{+a}+Q^+_a iv^{++}Q^{+a}\}~.
$$
The terms linear in $v^{++}$ and $q^+$ in (\ref{splitt}),
(\ref{hypeq}) determines the equation of motion and this term
should be dropped when considering the effective action.

To construct the effective action, we will follow the Faddeev-Popov
Ansatz. Within the framework of the background field method, we
should fix only the quantum transformation (\ref{qutr}). In
accordance with \cite{10}, let us introduce the gauge fixing
function in the form
$$
{\cal F}^{(4)}_\tau=D^{++}v^{++}_\tau =e^{-ib}({\cal
D}^{++}v^{++})e^{ib}=e^{-ib}{\cal F}^{(4)}e^{ib}~,
$$
which changes by the law
\begin{equation}\label{trf}
\delta{\cal F}^{(4)}_\tau=e^{-ib}\{{\cal D}^{++}({\cal
D}^{++}\lambda +i[v^{++},\lambda])\}e^{ib}
\end{equation}
under the quantum transformations (\ref{qutr}). Eg. (\ref{trf})
leads to the Faddeev-Popov determinant
$$\Delta_{FP}[v^{++},V^{++}]=\mbox{Det}({\cal D}^{++}({\cal D}^{++}
+iv^{++})).$$ To get a path-integral representation for
$\Delta_{FP}[v^{++},V^{++}]$, we introduce two real analytic
fermionic ghosts ${\bf b}$ and ${\bf c}$, in the adjoint
representation of the gauge group, and the corresponding ghost
action
\begin{equation}\label{FP}
S_{FP}[{\bf b}, {\bf c}, v^{++}, V^{++}] =\mbox{tr}\int
d\zeta^{(-4)}du {\bf b}{\cal D}^{++}({\cal D}^{++}{\bf c}
+i[v^{++},{\bf c}]).
\end{equation}
As a result, we arrive at the effective action $\Gamma[V^{++},
q^+]$ in the form
\begin{equation}\label{path}
e^{i\Gamma[V^{++},\; q^+]}=e^{iS_{cl}[V^{++},\; q^+]}\int {\cal
D}v^{++}{\cal D}Q^+ {\cal D}{\bf b}{\cal D}{\bf c}\times
\end{equation}
$$
\times e^{i(\Delta S_{SYM}[v^{++},V^{++}]+\Delta
S_H[v^{++},V^{++},Q^+,q^+]+S_{FP}[{\bf b}, {\bf c}, v^{++},
V^{++}])} \delta[{\cal F}^{(4)}-f^{(4)}]~,
$$
where $f^{(4)}(\zeta,u)$ is an external Lie-algebra valued analytic
superfield independent of $V^{++}$, and $\delta[{\cal F}^{(4)}]$ is
the proper functional analytic delta-function. To transform the path
integral for $\Gamma[V^{++},\; q^+]$ into a more useful form, we
average the right hand side in Eq. (\ref{path}) with the weight
\begin{equation}\label{weight}
\Delta[V^{++}]\exp\{\frac{i}{2\alpha}\mbox{tr}\int d^{12}z
du_1du_2
f_\tau^{(4)}(z,u_1)\frac{(u^-_1u^-_2)}{(u^+_1u^+_2)^3}f_\tau^{(4)}(z,u_2)\}.
\end{equation}
Here $\alpha$ is an arbitrary gauge parameter. The functional
$\Delta[V^{++}]$ should be chosen from the condition
\begin{equation}
1=\Delta[V^{++}]\int {\cal
D}f^{(4)}\exp\{\frac{i}{2\alpha}\mbox{tr}\int d^{12}z du_1du_2
f_\tau^{(4)}(z,u_1)\frac{(u^-_1u^-_2)}{(u^+_1u^+_2)^3}f_\tau^{(4)}(z,u_2)\}.
\end{equation}
Hence, using the standard identity $\int d\zeta^{(-4)}(D^+)^4
L(z,u) =\int d^{12}z L(z,u)$,
$$
\Delta^{-1}[V^{++}]=\int {\cal
D}f^{(4)}\exp\{\frac{i}{2\alpha}\mbox{tr}\int d\zeta^{(-4)}_1
d\zeta^{(-4)}_2 du_1du_2
f^{(4)}(\zeta_1,u_1)A(1,2)f^{(4)}(\zeta_2,u_2)\}
$$
$$
=\mbox{Det}^{-1/2}A~,
$$
we have the expression for $\Delta$ by means of a special
background-dependent operator\\
$A=\frac{(u^-_1u^-_2)}{(u^+_1u^+_2)^3}(D_1^+)^4(D_2^+)^4\delta^{12}(z_1-z_2)$
acting on the space of analytic superfields with values in the Lie
algebra of the gauge group. Thus
\begin{equation}\label{Det}
\Delta[V^{++}]=\mbox{Det}^{1/2}A~.
\end{equation}

To find $\mbox{Det} A$ we represent it by a functional integral
over analytic superfields of the form
\begin{equation}\label{Det1}
\mbox{Det}^{-1}A=\int {\cal D}\chi^{(4)}{\cal
D}\rho^{(4)}\exp\{i\mbox{tr}\int
d\zeta^{(-4)}_1du_1d\zeta^{(-4)}_2du_2 \chi^{(4)}(1)
A(1,2)\rho^{(4)}(2)\}~,
\end{equation}
and perform the following change of functional variables
$$
\rho^{(4)}=({\cal D}^{++})^2\sigma, \quad
\mbox{Det}(\frac{\delta\rho^{(4)}}{\delta\sigma})=\mbox{Det}({\cal
D}^{++})^2~.
$$
Then we have\footnote{we use
$(D^{++}_2)^2\frac{u^-_1u^-_2}{(u^+_1u^+_2)^3}=(u^+_2u^-_1)(D^{--}_2)^2\delta^{(3,-3)}(u_2,u_1)$}
\begin{equation}\label{Det2}
\mbox{tr}\int d\zeta^{(-4)}_1du_1d\zeta^{(-4)}_2du_2 \chi^{(4)}(1)
A(1,2)\rho^{(4)}(2)
\end{equation}
$$
=\mbox{tr}\int d^{12}z du_1du_2 \chi_\tau^{(4)}(1)
\frac{(u^-_1u^-_2)}{(u^+_1u^+_2)^3}(D^{++}_2)^2\sigma_\tau(2)
=\frac{1}{2}\mbox{tr}\int d^{12}z du \chi^{(4)}_\tau
(D^{--})^2\sigma_\tau
$$$$=-\mbox{tr}\int d\zeta^{(-4)}du
\chi^{(4)}{\stackrel{\frown}{\Box}}\sigma~,
$$
where
\begin{equation}\label{smile}
{\stackrel{\frown}{\Box}}=-\frac{1}{2}({\cal D}^+)^4({\cal
D}^{--})^2~.
\end{equation}
On the basis of Egs. (\ref{Det}-\ref{Det2}) one obtains
\begin{equation}
\Delta[V^{++}]=\mbox{Det}^{-1/2}({\cal D}^{++})^2
\mbox{Det}^{1/2}{\stackrel{\frown}{\Box}}_{(4,0)}~.
\end{equation}
Now, we are able to represent $\Delta[V^{++}]$ by the following
functional integral
\begin{equation}\label{NK}
\Delta[V^{++}]=\mbox{Det}^{1/2}{\stackrel{\frown}{\Box}}_{(4,0)}\int
{\cal D}\varphi e^{-\frac{i}{2}\mbox{tr}\int d\zeta^{(-4)}du {\cal
D}^{++}\varphi {\cal D}^{++}\varphi}~,
\end{equation}
with the integration variable $\varphi$ being a bosonic real
analytic superfield taking  values in the Lie algebra of the gauge
group. The $\varphi$ is, in fact, the Nielsen-Kallosh ghost for the
theory. As a result, we see that the ${\cal N}=2$ SYM theory is
described within the background field approach by three ghosts: the
two fermionic ghosts ${\bf b}$ and ${\bf c}$ and the third bosonic
ghost $\varphi$. The ghost action $S_{FP}$ and $S_{NK}$ given by
Eqs. (\ref{FP}) and (\ref{NK}) correspond to the known
$\omega$-multiplet \cite{6}, \cite{7}.

Upon averaging the effective action with the weight
(\ref{weight}), one gets the following path integral
representation
\begin{equation}\label{path2}
e^{i\Gamma[V^{++},\; q^+]}=e^{iS_{cl}[V^{++},\;
q^+]}\mbox{Det}^{1/2}{\stackrel{\frown}{\Box}}_{(4,0)}\end{equation}$$\times
\int {\cal D}v^{++}{\cal D}Q^+ {\cal D}{\bf b}{\cal D}{\bf c}{\cal
D}\varphi e^{iS_{q}[v^{++}, Q^+, {\bf b}, {\bf c}, \varphi,
V^{++}, q^+]}~,
$$
where
$$
S_{q}[v^{++}, Q^+, {\bf b}, {\bf c}, \varphi, V^{++}, q^+]=\Delta
S_{SYM}[v^{++, V^{++}}]+S_{GF}[v^{++}, V^{++}]$$$$ +\Delta
S_H[v^{++},V^{++},Q^+,q^+]+S_{FP}[{\bf b}, {\bf c}, v^{++},
V^{++}]+ S_{NK}[\varphi, V^{++}].
$$
Here $S_{GF}[v^{++}, V^{++}]$ is the gauge fixing contribution to
the quantum action
\begin{equation}\label{GF}
S_{GF}[v^{++}, V^{++}]=\frac{1}{2\alpha}\mbox{tr}\int d^{12}z du_1
du_2\frac{u^-_1u^-_2}{(u^+_1u^+_2)^3}(D^{++}_1
v^{++}_\tau(1)(D^{++}_2 v^{++}(2)))
\end{equation}
$$
=\frac{1}{2\alpha}\mbox{tr}\int d^{12}z du_1
du_2\frac{v_\tau^{++}(1)v_\tau^{++}(2)}{(u^+_1u^+_2)^2}-\frac{1}{4\alpha}\mbox{tr}\int
d^{12}z du v^{++}_\tau (D^{--})^2 v^{++}_\tau~.
$$
Let us consider the sum of the quadratic part in $v^{++}$ of
$\Delta S_{SYM}$ (\ref{splitt}) and $S_{GF}$ (\ref{GF}). It has
the form
$$
\frac{1}{2}(1+\frac{1}{\alpha})\mbox{tr}\int d^{12}z du_1
du_2\frac{v_\tau^{++}(1)v_\tau^{++}(2)}{(u^+_1u^+_2)^2}+\frac{1}{2\alpha}\mbox{tr}\int
d^{12}z du v^{++} {\stackrel{\frown}{\Box}} v^{++}~,
$$
where we have used Eq. (\ref{smile}). To further simplify the
computation, we make the simplest choice the Fermi-Feynman gauge
$\alpha=-1$. We can now write the final result for the effective
action $\Gamma[V^{++}, q^+]$
\begin{equation}\label{path3}
e^{i\Gamma[V^{++},\; q^+]}=e^{iS_{cl}[V^{++},\;
q^+]}\mbox{Det}^{1/2}{\stackrel{\frown}{\Box}}_{(4,0)} \int {\cal
D}v^{++}{\cal D}Q^+ {\cal D}{\bf b}{\cal D}{\bf c}{\cal D}\varphi
e^{iS_{q}[v^{++}, Q^+, {\bf b}, {\bf c}, \varphi, V^{++},
q^+]},\end{equation} where action $S_{q}$ is as follows
$$
S_{q}[v^{++}, Q^+, {\bf b}, {\bf c}, \varphi, V^{++},
q^+]=S_{2}[v^{++}, Q^+, {\bf b}, {\bf c}, \varphi, V^{++},
q^+]+S_{int}[v^{++}, Q^+, {\bf b}, {\bf c},  V^{++}, q^+],
$$
\begin{equation}\label{q2}
S_2=-\frac{1}{2}\mbox{tr}\int d\zeta^{(-4)}du
v^{++}{\stackrel{\frown}{\Box}} v^{++} +\mbox{tr}\int
d\zeta^{(-4)}du{\bf b}({\cal D}^{++})^2{\bf c}\end{equation}
$$+\frac{1}{2}\mbox{tr}\int d\zeta^{(-4)}du\varphi({\cal
D}^{++})^2\varphi +\frac{1}{2}\int d\zeta^{(-4)}du\{Q^+_a{\cal
D}^{++}Q^{+a}+Q^+_a iv^{++}q^{+a}+q^+_a iv^{++}Q^{+a}\}~,
$$
\begin{equation}\label{int}
S_{int}=-\mbox{tr}\int d^{12}z
du_1...du_n\sum_{n=3}^{\infty}\frac{(-i)^{n-2}}{n}\frac{v^{++}_\tau(z,u_1)...v^{++}_\tau(z,u_n)}{(u^+_1u^+_2)...(u^+_nu^+_1)}
\end{equation}
$$
-i\mbox{tr}\int d\zeta^{(-4)}du {\cal D}^{++}{\bf b}[v^{++}, {\bf
c}] +\frac{1}{2}\int d\zeta^{(-4)}du Q^+_a iv^{++}Q^{+a}~.
$$
This equations completely determine the structure of the
perturbation expansion for calculating the effective action
$\Gamma[V^{++}, q^+]$ of the ${\cal N}=2$ SYM theory with
hypermultiplets in a manifestly supersymmetric and gauge invariant
form. However not everybody hidden rigid symmetry of a classical
action can always be maintained manifestly in the Faddeev-Popov
quantization scheme. According to the analysis of \cite{Holt}, the
problem of keeping the rigid symmetries manifest at the quantum
level is essentially equivalent to finding covariant gauge
conditions. In the case of conformal symmetry, such gauge conditions
do not exist and any special conformal transformation has to be
accompanied by a field-dependent nonlocal gauge transformation in
order to restore the gauge slice \cite{defconf}, \cite{9}. The
invariance of the path integral under combined conformal and gauge
transformations lead to modified conformal Ward identities for the
effective action.

The action $S_2$ defines the propagators depending on background
fields \cite{10}. In the framework of the background field formalism
in ${\cal N}=2$ harmonic superspace there appear three types of
covariant matter and gauge field propagators. Associated with
$\stackrel{\frown}{\Box}$ is a Green's function $G^{(2,2)}(z,z')$
which is subject to the Feynman boundary conditions and satisfies
the equation $\stackrel{\frown}{\Box}G^{(2,2)}(1|2)=-{\bf
1}\delta^{(2,2)}(1|2)$, where the analytic
delta-function\footnote{$\delta^{(q,4-q)}(\zeta_1,u_1|\zeta_2,u_2)=(D^+_1)^4
\delta^{12}(z_1-z_2)\delta^{(q-4,4-q)}(u_1,u_2)=(D^+_2)^4
\delta^{12}(z_1-z_2)\delta^{(q,-q)}(u_1,u_2)$ \cite{6}.}
$\delta^{(2,2)}(\zeta_1,\zeta_2)$ is
$$
i<v^{++}(z,u)
v^{++}(z',u')>=G^{(2,2)}(z,u,z',u')=-\frac{1}{\stackrel{\frown}{\Box}}({\cal
D}^+)^4\{{\bf 1}\delta^{12}(z-z')\delta^{(-2,2)}(u,u')\}~.
$$
Sometimes, it is useful to rewrite $G^{(2,2)}$ in a manifestly
analytic at both points, following \cite{7}
\begin{equation}\label{provec}
G^{(2,2)}(1,2)=-\frac{1}{2\stackrel{\frown}{\Box}_1\stackrel{\frown}{\Box}_2}({\cal
D}^+_1)^4({\cal D}^+_2)^4\{{\bf
1}\delta^{12}(z_1-z_2)(D^{--}_2)^2\delta^{(-2,2)}(u_1,u_2)\}~.
\end{equation}
This representation may, in principle, be advantageous when handling
those supergraphs which contain a product of harmonic distributions.

The $Q^+$ hypermultiplet propagator associated with the action
(\ref{q2}) has the form
\begin{equation}\label{prohyp}
i<Q^+(\zeta_1,u_1, \zeta_2,u_2)>= G_b^{a (1.1)}(1| 2)=-\delta^a_b
\frac{({\cal D}^+_1)^4({\cal
D}^+_2)^4}{(u^+_1u^+_2)^3}\frac{1}{{\stackrel{\frown}{\Box}}_1}
\delta^{12}(z_1-z_2)~.
\end{equation}
It is not hard to see that this manifestly analytic expression is
the solution of the equation ${\cal
D}^{++}_1G^{(1,1)}=\delta_A^{(3,1)}(1|2).$ For the hypermultiplet of
the second type described by a chargeless real analytic superfield
$\omega(\zeta,u)$ the equation for Green' function is $({\cal
D}^{++}_1)^2G^{(0,0)}(1|2)=\delta_A^{(4,0)}(1|2)$. The suitable
expression for $G^{(0,0)}$ is
\begin{equation}\label{proomeg}
i<\omega(1),
\omega^T(2)>=G^{(0,0)}(1|2)=-\frac{1}{{\stackrel{\frown}{\Box}}_1}({\cal
D}^+_1)^4({\cal D}^+_2)^4\{{\bf
1}\delta^{12}(z_1-z_2)\frac{u^-_1u^-_2}{(u^+_1u^+_2)^3}\}.
\end{equation}
Switching off the gauge background superfield, the Green's
functions (\ref{provec}, \ref{prohyp}, \ref{proomeg}) turn into
the free ones obtained in \cite{6}, \cite{7}. The operator
${\stackrel{\frown}{\Box}}=-\frac{1}{2}({\cal D}^+)^4({\cal
D}^{--})^2$ transforms each covariantly analytic superfield into a
covariantly analytic and, using algebra (\ref{alg}), can be
rewritten as second-order d'Alemberian-like differential operator
on the space of such superfields \cite{10}:
\begin{equation}\label{smile}
\stackrel{\frown}{\Box} =\frac{1}{2}{\cal
D}^{\alpha\dot\alpha}{\cal D}_{\alpha\dot\alpha}
+\frac{i}{2}({\cal D}^{+\alpha}{\cal W}){\cal D}^-_\alpha
+\frac{i}{2}(\bar{\cal D}^{+}_{\dot\alpha}\bar{\cal W})\bar{\cal
D}^{-\dot\alpha} +\frac{1}{2}\{{\cal W}, \bar{\cal W}\}
\end{equation}
$$
-\frac{i}{4}(\bar{\cal D}^+\bar{\cal D}^+\bar{\cal W}){\cal
D}^{--} +\frac{i}{8}[{\cal D}^+,{\cal D}^-]{\cal W}~,
$$
Among the important properties of $\stackrel{\frown}{\Box}$ is the
following: $({\cal
D}^+)^4\stackrel{\frown}{\Box}=\stackrel{\frown}{\Box}({\cal
D}^+)^4.$ The coefficients of this operator depend on background
super\-fields ${\cal W}, \bar{\cal W}$. For the background,
belonging to an Abelian subgroup of the gauge group and satisfying
the on-shell conditions, we have the further restriction: ${\cal
D}^{\pm}{\cal W}=D^{\pm}{\cal W}$, and similarly for $\bar{\cal W}$
with $D,\bar{D}$ being background independent derivatives. In
addition, we should omit the two last terms in (\ref{smile}) since
they disappear on-shell.

To sim\-plify the quad\-ra\-tic part of the action for quantum
gauge superfields it is convenient to expand these superfields in
some basis. We choose the quantum superfields in one-to-one
correspondence with the roots of the Lie algebra of gauge group
$G$: $v=\sum_\alpha v^\alpha {E}_\alpha +\sum_i v^i {H}_i$. Here
${ E}_\alpha$ is the generator corresponding to the root $\alpha$
normalized as $\mbox{tr}({ E}_\alpha {
E}_{-\beta})=\delta_{\alpha,-\beta}$ and ${H}_i$ are the
rank$({G})$ Cartan subalgebra generators satisfying the
commutation relations $[{H}_i, {E}_\alpha] =\alpha({H}_i) {
E}_\alpha$. Using these notations one can rewrite the actions
(\ref{q2}) in terms of coefficients in expansions of $v$ in above
basis. Such form of writing of the effective action is very
convenient for its evaluation and  will be used in Section 4 for
various cases.

\section{Structure of the one-loop effective action}
Consider the loop expansion of the effective action within the
background field formulation. Then, the effective action is given
by vacuum diagrams (that is diagrams without external lines) with
background field dependent propagators and vertices. A formal
expression of the one-loop effective action $\Gamma[V^{++}, \;
q^+]$ for the theory under consideration is written in terms of a
path integral as follows \cite{10}:
\begin{equation}\label{def}
e^{i\Gamma[V^{++}, \; q^+]}=e^{iS(V^{++},\;
q^+)}\mbox{Det}^{1/2}\stackrel{\frown}{\Box}_{(4.0)}\int {\cal D}
v^{++}{\cal D} {\bf b} {\cal D} {\bf c} {\cal D}\varphi {\cal D}
Q^+ \; e^{iS_2[v^{++}, {\bf b}, {\bf c}, \varphi, Q^+, V^{++},
q^+]}~,
\end{equation}
where the full quadratic action is defined in Eq. (\ref{q2}):
\begin{equation}\label{s2}
S^{(2)}[v^{++}, {\bf b}, {\bf c},\varphi, Q^+, V^{++},
q^+]=-\frac{1}{2}\mbox{tr} \int d\zeta^{(-4)}du\, v^{++}
\stackrel{\frown}{\Box} v^{++} +\mbox{tr} \int d\zeta^{(-4)}du
\,{\bf b} ({\cal D}^{++})^2{\bf c}
\end{equation}
$$+
\frac{1}{2}\mbox{tr}\int d\zeta^{(-4)}du \,\varphi ({\cal
D}^{++})^2\varphi-\frac{1}{2}\int d\zeta^{(-4)} (Q^{a+}{\cal
D}^{++}Q^+_a +q^{+a}iv^{++}Q^+_a +Q^{+a}iv^{++}q^+_a)~.
$$
Here $v^{++}$ is a quantum vector superfield taking values in the
Lie algebra of the gauge group and ${\bf b}$, ${\bf c}$ are two real
analytic Faddeev-Popov fermionic ghosts and $\varphi$ is the bosonic
Nielsen-Kallosh ghost, all in the adjoint representation of the
gauge group. Eqs.(\ref{def}), (\ref{s2}) completely determine the
structure of the perturbation expansion for calculating the
effective action of the ${\cal N}=2$ SYM with hypermultiplets in a
manifestly supersymmetric and gauge invariant form. For the
propagators of the quantum vector multiplet $v^{++}$ and the
hypermultiplets $Q^{+a}$ we use (\ref{provec}) and (\ref{prohyp})
respectively. Vertices can be taken directly from the second line
(\ref{s2}). It is easy to see that the ghosts do not couple to
background hypermultiplet and therefore do not contribute to
hypermultiplet dependent part of the one-loop effective action. In
the vector sector of the ${\cal N}=2$ SYM theory where the matter
hypermultiplet are integrated out, the one-loop effective action
$\Gamma [V^{++}]$ reads
$$
\Gamma
[V^{++}]=\frac{i}{2}\mbox{Tr}_{(2,2)}\ln\stackrel{\frown}{\Box}
-\frac{i}{2}\mbox{Tr}_{(4,0)}\ln\stackrel{\frown}{\Box}-\frac{i}{2}\mbox{Tr}_{ad}\ln({\cal
D}^{++})^2 +i\mbox{Tr}_{R_q}\ln{\cal D}^{++}
+\frac{i}{2}\mbox{Tr}_{R_\omega}\ln({\cal D}^{++})^2.
$$
Currently, the holomorphic and non-holomorphic parts of the
low-energy effective action ${\cal N}=2,4$ SYM theory on the Coulomb
branch, including Heisenberg-Euler type action in the presence of a
covariantly constant vector multiplet, are completely known (see for
a review e.g. \cite{n2}, \cite{n4}). The general structure of the
low-energy effective action in ${\cal N}=2,4$ superconformal
theories is \cite{bkt}:
$$
\Gamma=S_{cl}+\int d^{12}z\{c\ln{\cal W}\ln\bar{\cal W}+\int
d^{12}z\ln{\cal W}\Lambda(\frac{D^4\ln{\cal W}}{\bar{\cal
W}^2})+c.c.$$$$\quad \quad \quad \quad \quad \quad \quad \quad
\quad \quad \quad \quad +\int
d^{12}z\Upsilon(\frac{\bar{D}^4\ln\bar{\cal W}}{{\cal
W}^2},\frac{D^4\ln{\cal W}}{\bar{\cal W}^2})\}+...~,
$$
where $\Lambda$ and $\Upsilon$ are holomorphic and real analytic
function of the (anti)chiral superconformal invariants. The $c$-term
is known to generate four-derivative quantum corrections at the
component level which include an famous $F^4$ term (see e.g.
\cite{n4}).The hypermultiplet dependent part of the effective action
in ${\cal N}=4$ SYM theory in leading order is also known \cite{bi}
- \cite{bp}.

For further analysis of the effective action it is convenient to
diagonalize the action of quantum fields $S^{(2)}$ using a special
shift of hypermultiplet variables in the path integral\footnote{In
QED such a change of variables has been fulfilled in ref. \cite{11}.
N.G.P is grateful to S. Kuzenko for bringing his attention to this
reference.}
\begin{equation}\label{replac}
Q^{+a}= \xi^{+a} + i\int d \zeta^{(-4)}_2  q^{+b}(2)v^{++}(2) G_b^{a
(1.1)}(1|2)~,
\end{equation}
$$
Q^{+}_a= \xi^{+}_a - i\int d \zeta^{(-4)}_2 G_a^{b (1.1)}(1|2)
 v^{++}(2)q^{+}_{b}(2)~,
$$
where $\xi^{+a}, \xi^{+}_{a}$ are the new independent variables in
the path integral. It is evident that the Jacobian of the
replacement (\ref{replac}) is equal to unity. Here $G_b^{a (1.1)}(1|
2)$ is the background-dependent propagator (\ref{prohyp}) for the
superfields $Q^{+a}, Q^+_b$. In terms of the new set of quantum
fields we obtain for the following hypermultiplet dependent part of
the quadratic action
$$
S^{(2)}_H=-\frac{1}{2}\int d\zeta^{(-4)} \xi^{a+}{\cal
D}^{++}\xi^+_a -\frac{1}{2}\int d\zeta^{(-4)}_1 d\zeta^{(-4)}_2
q^{+a}(1)v^{++}(1) G_a^{b (1.1)}(1|2)v^{++}(2)q^+_b(2)~.
$$
Then the vector multiplet dependent part of the quadratic action
gets the following non-local extension
\begin{equation}\label{green2}
S^{(2)}_v=-\frac{1}{2}\mbox{tr}\int d \zeta^{(-4)}_1 v^{++}_1 \int d
\zeta^{(-4)}_2\left( {\stackrel{\frown}{\Box}}\delta^{(2.2)}_A(1|2)
+q^{+a}(1)G_a^{b(1.1)}(1|2)q^+_b(2) \right) v^{++}_2~.
\end{equation}
Expression (\ref{green2}), written as an analytical nonlocal
superfunctional, will be a starting point for our calculations of
the one-loop effective action in the hypermultiplet sector. Our aim
in the current and later sections is to find the leading low-energy
contribution to the effective action for the slowly varying
hypermultiplet when all derivatives of the background hypermultiplet
can be neglected. We will show that for such a case the non-local
interaction is localized.

Using the relation $v^{++}_2=\int d\zeta_3^{(-4)} \delta^{(2.2)}_A
(2|3)v^{++}_3$ one can rewrite expression for $S^{(2)}_v$
(\ref{green2}) in the form
\begin{equation}\label{nonlocal}
S^{(2)}_v=-\frac{1}{2}\mbox{tr}\int d \zeta^{(-4)}_1\, v^{++}_1 \int
d \zeta^{(-4)}_2\left(
{\stackrel{\frown}{\Box}}\delta^{(2.2)}_A(1|2)\right.
\end{equation}
$$\left.\quad \quad \quad \quad \quad \quad \quad \quad +\int
d\zeta^{(-4)}_3
q^{+a}(1)G_a^{b(1.1)}(1|3)q^+_b(3)\delta^{(2.2)}_A(3|2)
\right)v^{++}_2~.
$$
Then we use the explicit form of the Green function (\ref{prohyp})
and the relation allowing us to express the $({\cal
D}^{+}_1)^4({\cal D}^+_2)^4$ as a polynomial in powers of
$(u^+_1u^+_2)$ \cite{{skim}}
\begin{equation}\label{polin}
\begin{array}{c} ({\cal D}^+_1)^4({\cal D}^+_2)^4=\\
({\cal D}^+_1)^4\left(({\cal D}^-_1)^4(u^+_1u^+_2)^4
-\frac{i}{2}\Delta_1^{--}(u^+_1u^+_2)^3(u^-_1u^+_2)
-{\stackrel{\frown}{\Box}}_1(u^+_1u^+_2)^2(u^-_1u^+_2)^2\right)~,
\end{array}
\end{equation}
where the operator $\Delta^{--}$ is \cite{{skim}}
\begin{equation}\label{Delt}
\Delta^{--}={\cal D}^{\alpha\dot\alpha}{\cal D}^-_\alpha \bar{\cal
D}^-_{\dot\alpha} +\frac{1}{2}{\cal W}({\cal D}^-)^2
+\frac{1}{2}\bar{\cal W}(\bar{\cal D}^-)^2 +({\cal D}^-{\cal
W}){\cal D}^- +(\bar{\cal D}^-\bar{\cal W})\bar{\cal D}^-~.
\end{equation}
Since $G^{(1.1)}(1,2)=-G^{(1.1)}(2,1),$  the non-local term in
(\ref{nonlocal}) takes the form
$$
\int d\zeta^{(-4)}_3 q^{+a}(1)({\cal D}^+_3)^4\left(({\cal
D}^-_3)^4(u^+_3u^+_1)\frac{1}{{\stackrel{\frown}{\Box}}_3}\right.$$
$$
\left.-\frac{i}{2}\Delta^{--}_3(u^-_3u^+_1)\frac{1}{{\stackrel{\frown}{\Box}}_3}
-\frac{(u^-_3u^+_1)^2}{(u^+_3u^+_1)}\right)\delta^{12}(1|3)q^+_a(3)\delta^{(2.2)}_A(3|2)~.
$$
The large braces here contain three terms. It is easy to see that
two first terms include the derivatives which will lead to
derivatives of the hypermultiplet in the effective action. Since we
keep only contributions without derivatives, the above terms can be
neglected. As a result, is it sufficient to consider only the third
term in the braces.

Now we apply the relation $\int d\zeta_3^{(-4)}({\cal D}^+_3)^4=\int
d^{12}z,$ allowing to integrate over $z_{3}$, and obtain
$$-\int du_3\,
q^{+a}(1)\frac{(u_3^-u^+_1)^2}{(u^+_3u^+_1)}q^+_a(u_3, z_1)
\delta^{(2.2)}_A(u_3,z_1|2)~.
$$
Then one uses the on-shell harmonic dependence of hypermultiplet
$q^{+a}(3)=u^+_{3i}q^{ia}$ and take the coincident limit $u_1=u_3$
(conditioned by $\delta^{(2.2)}_A(u_3,z_1|2)$). After that we get
$\int du_3 \frac{u^+_{3 i}}{u^+_3u^+_1}=-u^-_{1 i}$. As a result,
the term under consideration has the form
$$q^{+a}(1)q^-_a(1)\delta^{(2.2)}_A(1|2),$$ where the expression
$q^{+a}(1)q^-_a(1)= q^{ia}q_{ia}$ is treated further as the slowly
varying superfield and all its derivatives are neglected. Namely
such an expression was obtained in \cite{bp} by summation of
harmonic supergraphs.

Thus, the second term in (\ref{nonlocal}) becomes local in the
leading low-energy approximation. As a result, the operator in
action $S^{(2)}_v$ determining the effective background covariant
propagator of the quantum vector multiplet superfield $v^{++}_I$ (we
expanded the superfield $v^{++} $ in generators $v^{++}=v^{++}_I
T_I$ and work further only with superfield components $v^{++}_I$)
takes the form
\begin{equation}\label{kin}
\left(\stackrel{\frown}{\Box}_{IJ}
+q^{+a}(z_1,u_1)\{T_I,T_J\}q^-_a(z_1,u_1)\right)\delta^{(2.2)}_A(1|2)~,
\end{equation}
where
$$
\stackrel{\frown}{\Box}_{IJ} = \mbox{tr}\left(T_{(I}{\Box}T_{J)}
+\frac{i}{2}T_{(I}[{\cal D}^{+\alpha}{\cal W},T_{J)}]{\cal
D}^-_\alpha +\frac{i}{2}T_{(I}[\bar{\cal
D}^+_{\dot\alpha}\bar{\cal W},T_{J)}]\bar{\cal D}^{-\dot\alpha}
+T_{(I}[{\cal W},[\bar{\cal W}, T_{J)}]]\right)~.
$$
Here ${\Box} = \frac{1}{2}{\cal D}^{\alpha\dot\alpha}{\cal
D}_{\alpha\dot\alpha}$ is the covariant d'Alemberian.

Thus, using the ${\cal N}=2$ harmonic superspace formulation of the
${\cal N}=2$ SYM theory with hypermultiplets and techniques of the
non-local shift we obtained that the whole dependence on the
background hypermultiplet is concentrated in the quantum vector
multiplet sector with the modified quadratic action. Therefore the
one-loop effective action is given by the expression
\begin{equation}\label{gamma}
\Gamma^{(1)}[V^{++}, q^+] = \Gamma^{(1)}_v[V^{++}, q^+] +
\widetilde{\Gamma}^{(1)}[V^{++}]~,
\end{equation}
where the first term in (\ref{gamma}) is originated from quantum
vector multiplet $v^{++}_I$
\begin{equation}\label{gamma1}
\Gamma^{(1)}_v[V^{++}, q^+]=\frac{i}{2}\mbox{Tr}\ln(
\stackrel{\frown}{\Box}_{IJ} +q^{+a }\{T_I,T_J\}q^-_{a})~.
\end{equation}
Second term in (\ref{gamma}) is the contribution of ghosts and
quantum hypermultiplet $\xi^{+}_a$ and does not depend on the
background hypermultiplet.

As a result, the background hypermultiplet dependence of one-loop
effective action is included into the operator
\begin{equation}\label{operator}
\stackrel{\frown}{\Box}_{IJ} + q^{+a}\{T_I,T_J\}q^{-}_{a}~,
\end{equation}
acting on $v^{++}_I$ and containing the mass matrix of the vector
multiplet
\begin{equation}\label{mass1}
({\cal M}^{2}_{v})_{IJ} = \mbox{tr}\left([T_I,{\cal W}][\bar{\cal
W},T_J]+(I\leftrightarrow J)\right) + q^{+a}\{T_I,T_J\}q^{-}_{a}~,
\end{equation}
if $q^+$ is in the fundamental representation, and
\begin{equation}\label{mass2}
({\cal M}^{2}_{v})_{IJ} = \mbox{tr}\left([T_I,{\cal W}][\bar{\cal
W},T_J] + [q^{+a},T_I][T_J,q^{-}_{a}]\right) +(I\leftrightarrow
J)~,
\end{equation}
if $q^+$ in an arbitrary matrix representation. We have proved that
the hypermultiplet dependence is completely transferred into the
sector of quantum superfields $v^{++}$ and conditioned by the
background covariant operator (\ref{operator}). Eqs. (\ref{gamma}),
(\ref{gamma1}) are a starting point for calculating the one-loop
effective action. Note that we make no restrictions on a space-time
dependence of the hypermultiplet except the on-shell properties
(\ref{onsh}).

In the above discussion, the gauge group structure of the
superfields ${\cal W}, q^+_a$ has been completely arbitrary.
Henceforth, the background superfields will be chosen to be aligned
along a fixed direction in the moduli space of vacua in such a way
that their scalar fields should solve Egs. (\ref{vacua}). Let the
background vector multiplet and hypermultiplet be of the form
(\ref{vac}) where ${H}$ is a fixed generator in the Cartan
subalgebra. It corresponds to assumption that gauge group ${G}$ is
broken down to ${\tilde{G}} \times K$ where $K$ is an Abelian
subgroup conditioned by the Cartan subalgebra where the generator
${H}$ belongs to. In this case there is a single vacuum combination
${\cal W}\bar{\cal W}$ for ${\cal N}=2$ background vector multiplet
and a single vacuum combination $q^{+a}q^{-}_{a}$ for the background
hypermultiplet\footnote{If the background fields corresponds to
various generators $H_{i}$ the effective action will be a sum of
contributions over the index $i$ where each contribution has the
structure corresponding to the above case. Therefore we can consider
the case with fixed generator $H$ without loss of generality. }.
Then the operator acting on the quantum vector multiplet superfields
defined in (\ref{operator}) takes the universal form
\begin{equation}\label{univer}
\Box +\frac{i}{2}\alpha({H})({\cal D}^+{\cal W}{\cal D}^-
+\bar{\cal D}^+\bar{\cal W}\bar{\cal D}^- ) + \alpha^2({ H}){\cal
W}\bar{\cal W} + q^{+a}q^{-}_{a}Z~,
\end{equation}
where $\Box$ is the covariant d'Alemberian, the combination
$q^{+a}q^{-}_{a}$ ($a=1,2$) already has no matrix indices (since a
fixed direction in moduli space is taken) and the matrix $Z$ has
the indices $I,J$ conditioned by the expression $\{T_I, T_J\}$
after fixation of the background hypermultiplet in accordance with
(\ref{vac}). All matrices containing ${\cal W}, \bar{\cal W}$ in
(\ref{univer}) are diagonal over the indices of the generators in
${\tilde{G}}$.

We are interesting only in the hypermultiplet dependent terms in the
one-loop effective action (\ref{gamma1}). Let us clarify how such
terms can in principle be generated in (\ref{gamma1}). The mass
matrix has the structure ${\cal M}^{(2)}_{v} = \alpha^2({H}){\cal
W}\bar{\cal W}\cdot Y + q^{+a}q^{-}_{a}\cdot Z$. The only eigenvalue
of the matrix $Y$ is $1$ with $n(H)$ corresponding eigenvectors. The
matrix in parentheses in (\ref{univer}) has the same eigenvectors as
$Y$. As to the matrix $Z$, there can be two options:

$({\bf i})$ The matrix $Z$ has $n(\Upsilon)$ eigenvectors common
with eigenvectors of $Y$ ($n(\Upsilon)\le n(H)$) and the
corresponding eigenvalues are $r({\Upsilon})$. Then the effective
action is the sum over different values of $r({\Upsilon})$.
Therefore we assume, without loss of generality, that there is only
one eigenvalue $r({\Upsilon})$ with $n(\Upsilon)$ eigenvectors
common with eigenvectors of Y. Hence the hypermultiplet dependent
effective action in the case under consideration is
\begin{equation}\label{gamma2}
\Gamma^{(1)}_v[V^{++},
q^+]=\end{equation}$$\frac{i}{2}n(\Upsilon)\mbox{Tr}\ln\left( \Box
+ \frac{i}{2}\alpha({H})({\cal D}^+{\cal W}{\cal D}^- +\bar{\cal
D}^+\bar{\cal W}\bar{\cal D}^- ) + \alpha^2({H}) {\cal W}\bar{\cal
W} + r(\Upsilon)q^{+a}q^{-}_{a} \right)~.
$$
Here $\mbox{Tr}$ means the functional trace of operators acting on
analytic superfields of the appropriate $U(1)$
charge\footnote{Specifically, if ${\cal
A}^{(p,\,4-p)}(\zeta_1,\zeta_2)$ is the kernel of an operator acting
on space of covariantly analytic superfields of charge $p$, then
$$
\mbox{Tr} {\cal A}^{(p,\,4-p)}=\mbox{tr}\int d\zeta^{(-4)}du {\cal
A}^{(p,\,4-p)}(\zeta, \zeta),
$$
where the trace $'\mbox{tr}'$is over group indices.}. The
eigenvectors of $Y$ which do not coincide with eigenvectors of $Z$
and give no hypermultiplet dependent contributions to the
effective action.

$({\bf ii})$ The matrices $Y$ and $Z$ have no common eigenvectors.
The effective action is the sum over eigenvectors of $Y$ and
eigenvectors of $Z$. The contribution originated from the
eigenvectors of $Y$ are hypermultiplet independent. The
contributions originated from the eigenvectors of $Z$ do not contain
the operators ${\cal D}^-$ and $\bar{\cal D}^-$ since the
corresponding matrix in (\ref{operator}) has different eigenvectors
then $Y$. However these operators are used to obtain, in principle,
the non-zero low-energy effective action. Therefore in this case the
hypermultiplet dependent part of the effective action vanishes.

As the result, the hypermultiplet dependent effective action is
given by the expression (\ref{gamma2}). In the next section we will
consider the evaluation of this expression.

\section{Calculation of the one-loop effective action}
The expression (\ref{gamma2}) is a basis for an analysis of the
hypermultiplet dependence of the effective action. This expression
will be written in the convenient form allowing us to evaluate it
using the superfield proper time techniques (see \cite{book},
\cite{other} for ${\cal N}=1$ superfield proper time techniques) by
generalizing the Schwinger construction \cite{schwinger}. We will
follow the generic approach developed in our paper \cite{bp} where
hypermultiplet dependence of ${\cal N}=4$ SYM effective action was
analyzed.

In the framework of the  Fock - Schwinger proper-time
representation, the effective action (\ref{gamma2}) is written as
follows
\begin{equation}\label{proper}
\Gamma^{(1)}_v [V^{++}, q^+]=\frac{i}{2} n(\Upsilon) \int d
\zeta^{(-4)} du \int^\infty_0 \frac{ds}{s}e^{-s(\Box
+\frac{i}{2}\alpha({H})({\cal D}^+{\cal W}{\cal D}^- +\bar{\cal
D}^+\bar{\cal W}\bar{\cal D}^-) + {\cal M}^2_v)}\times
\end{equation}
$$
\times ({\cal D}^+)^4
\left(\delta^{12}(z-z')\delta^{(-2,2)}(u,u')\right)|_{z=z',u=u'}=\int_0^\infty
\frac{d s}{s} \mbox{Tr} K(s),
$$
where ${\cal M}^2_v=\alpha^2({H}){\cal W}\bar{\cal W}+ r(\Upsilon)
q^{+a}q^-_a$. Here $K(s)$ is a superfield heat kernel, the operation
$\mbox{Tr}$ means the functional trace in the analytic subspace of
the harmonic superspace $\mbox{Tr}K(s)=\mbox{tr}\int
d\zeta^{(-4)}K(\zeta,\zeta|s)$, where $\mbox{tr}$ denotes the trace
over the discrete indices.  Representation of the effective action
(\ref{proper}) allows us to develop a straightforward evaluation of
the effective action in a form of covariant spinor derivatives
expansion in the superfield Abelian strengths ${\cal W}, \bar{\cal
W}$. The leading low-energy terms in this expansion correspond to
the constant space-time background $D^-_\alpha D^+_\beta {\cal
W}=\mbox{const}$, $\bar{D}^-_{\dot\alpha} \bar{D}^+_{\dot\beta}
\bar{\cal W}=\mbox{const}$ and on-shell background hypermultiplet.
Especially we want to emphasize that on-shell conditions do not mean
that the hypermultiplet is constant. Furthermore we assume that
hypermultiplet is a slowly varying function in the superspace and
neglect any derivatives of the hypermultiplet for deriving  the
superfield effective action. However, it does not mean that we miss
all space-time derivatives in the component effective Lagrangian.
Grassmann measure in the integral over harmonic superspace
$d^4\theta^{+}d^4\theta^{-}$ generates four space-time derivatives
in component expansion of the superfield Lagrangian. Therefore the
above assumption is sufficient to obtain  a component effective
Lagrangian including four space-time derivatives of the scalar
components of the hypermultiplet. Possible contributions to the
hypermultiplet dependent effective action off-shell will be
discussed in the next section.

Calculation of the effective action (\ref{proper}) is based on
evaluating the superfield heat kernel $K(s)$.
Note that even with regard to the properties of a non-analytic
integrand, the crucial idea \cite{km} is to stay in the analytic
subspace at all stages of the calculations without an artificial
conversion of the analytic integral into the full superspace
integral, where, as a rule, an integrand contains ill-defined
products of harmonic distributions. In this respect, integration
with the analytic measure will be a high-power projector which
removes all harmonic singularities. Moreover, because the covariant
d'Alemberian doesn't contain efficiently acting ${\cal D}^+$ we,
during heat kernel treatment, never obtain ${\cal D}^+q^-$ but only
${\cal D}^-q^-=0$. Under this approach we should actually look for
higher-derivative quantum corrections in the form \begin{equation}
\int d\zeta^{(-4)}({\cal D}^+)^4 {\cal H}({\cal W}, \bar{\cal W},
q^+, q^-). \end{equation}

In the case of a covariantly constant hypermultiplet ${\cal D}_m
q^+=0$ and vector multiplet ${\cal D}_m {\cal W}={\cal D}_m
\bar{\cal W}=0$ the heat kernel can be computed exactly. In order
to obtain the complete kernel it is convenient separate of the
contributions of the diamagnetic and paramagnetic parts of the
operator $\stackrel{\frown}{\Box}$. We follow here a generic
scheme of calculations \cite{bp} taking into account only the
aspects essential for the theory under consideration. As the first
step we use the Baker-Campbell-Hausdorff relation and write the
operator $K(s)$ as a products of several operator exponents
\footnote{Where we have used the notations
$$ A^{+\alpha}=\frac{i}{2}\alpha(H)({\cal D}^{+\alpha}{\cal W}), \quad
\bar{A}^{+\dot\alpha}=-\frac{i}{2}\alpha(H)(\bar{\cal
D}^{+\dot\alpha}\bar{\cal W}),\quad {\cal N}_\alpha^\beta
=D^-_\alpha A^{+\beta}, \quad \bar{\cal
N}_{\dot\alpha}^{\dot\beta} =\bar{D}^-_{\dot\alpha}
\bar{A}^{+\dot\beta}
$$}
\begin{equation}\label{kern}
{K}(s)=\exp(-s\{A^+{\cal D}^- + \bar{A}^+\bar{\cal D}^-
+\frac{1}{2}{\cal D}^{\alpha\dot\alpha}{\cal
D}_{\alpha\dot\alpha}+{\cal M}^2_v\})
\end{equation}
$$
= \exp\{-f_{\alpha\dot\alpha}(s){\cal
D}^{\alpha\dot\alpha}\}\exp\{-s\frac{1}{2}{\cal
D}^{\alpha\dot\alpha}{\cal
D}_{\alpha\dot\alpha}\}\exp\{-\Omega(s)\}\exp\{-s(A^+{\cal D}^- +
\bar{A}^+\bar{\cal D}^- )\}~.
$$
with some unknown coefficients in the right hand side. These
coefficients can by found directly by solving the system of a
differential equations on these coefficients. To find the mentioned
system of equations we consider $(\frac{d}{ds}K(s))K^{-1}(s)$ and
substitute for $K(s)$ first and second lines in (\ref{kern})
subsequently. Equations for the function $f^{\dot\alpha\alpha}(s)$
have the form
$$
\frac{d}{ds}f_{\alpha\dot\alpha}(s)=-f_{\beta\dot\beta}F^{\dot\beta\beta}_{\dot\alpha\alpha}-A^{+\beta}(D^-_\beta
f_{\alpha\dot\alpha})-\bar{A}^{+\dot\beta}(\bar{D}^-_{\dot\beta}
f_{\alpha\dot\alpha})
$$
$$
\quad \quad \quad \quad +A^+_\beta \bar{A}^-_{\dot\beta}(\int_0^s
d\tau e^{\tau F})^{\dot\beta\beta}_{\dot\alpha\alpha}
+\bar{A}^+_{\dot\beta} {A}^-_{\beta}(\int_0^s d\tau e^{\tau
F})^{\dot\beta\beta}_{\dot\alpha\alpha}~.
$$
It is easy to show that the solution  of these equation can be
written as follows
\begin{equation}
f_{\alpha\dot\alpha}= -A^+_\delta {\cal
F}^{\delta\dot\delta}_{\alpha\dot\alpha}\bar{A}^-_{\dot\delta}
-\bar{A}^+_{\dot\delta}\bar{\cal
F}^{\delta\dot\delta}_{\alpha\dot\alpha}A^-_\delta,
\end{equation}
where the function ${\cal F}({\cal N},\bar{\cal N},s), \bar{\cal
F}({\cal N},\bar{\cal N},s)$ are listed in \cite{bp}. Analogously,
equation for the function $\Omega$ is
$$
\frac{d}{ds}\Omega(s)-{\cal
M}^2_v=-A^{+\alpha}(D^-_\alpha\Omega)-\bar{A}^{+\dot\alpha}(\bar{D}^-_{\dot\alpha}\Omega)+A^+_\alpha
f^{\alpha\dot\alpha}\bar{A}^-_{\dot\alpha}+\bar{A}^+_{\dot\alpha}
f^{\dot\alpha\alpha}A^-_{\alpha}
$$
$$
-\frac{1}{2}A^+_\beta \bar{A}^-_{\dot\beta}(\int^s_0 d\tau
e^{-\tau
F})^{\dot\beta\beta}_{\dot\alpha\alpha}F^{\dot\alpha\alpha}_{\dot\rho\rho}f^{\dot\rho\rho}-\frac{1}{2}\bar{A}^+_{\dot\beta}
A^-_{\beta}(\int^s_0 d\tau e^{-\tau
F})^{\dot\beta\beta}_{\dot\alpha\alpha}F^{\dot\alpha\alpha}_{\dot\rho\rho}f^{\dot\rho\rho}.
$$
Solution of these equation has the form
\begin{equation}
\Omega(s)=s{\cal M}^2_v +A^{+\alpha}\Omega^-_\alpha(s)
+\bar{A}^{+\dot\alpha}\bar\Omega^-_{\dot\alpha}(s)+(A^+)^2\Psi^{(-2)}(s)+(\bar{A}^+)^2\bar\Psi^{(-2)}(s)
\end{equation}
$$
+A^{+\alpha}\bar{A}^+_{\dot\alpha}\Psi^{\dot\alpha(-2)}_\alpha(s).
$$
We point out that this solution is a finite order polynomials in
power of the Grassmannian elements $A^{\pm}, \bar{A}^{\pm}.$ All
coefficients  are given in ref. \cite{bp}. Now it is necessary to
write the last exponential in (\ref{kern}) in the form
\begin{equation}
\exp\{-s(A^+{\cal D}^- + \bar{A}^+\bar{\cal
D}^-)\}=1+a^{+\alpha}{\cal D}^-_\alpha
+\bar{a}^{+\dot\alpha}\bar{\cal D}^-_{\dot\alpha}+f^{+2}({\cal
D}^-)^2+\bar{f}^{+2}(\bar{\cal D}^-)^2
\end{equation}
$$
+f^{+2\dot\alpha\alpha}{\cal D}_\alpha \bar{\cal
D}_{\dot\alpha}+\bar\Xi^{+3\dot\alpha}\bar{\cal
D}^-_{\dot\alpha}({\cal D}^-)^2 +\Xi^{+3\alpha}{\cal
D}^-_{\alpha}(\bar{\cal D}^-)^2+\Omega^{+4}({\cal
D}^-)^2(\bar{\cal D}^-)^2.
$$
The coefficients of this expansion can be found exactly and given in
\cite{bp}. For further analysis it is important to note that
\begin{equation}
\Omega^{+4}=-\frac{1}{16}(A^+)^2(\bar{A}^+)^2\mbox{tr}\left(\frac{\cosh(s{\cal
N})-1}{{\cal N}^2}\right)\mbox{tr}\left(\frac{\cosh(s\bar{\cal
N})-1}{\bar{\cal N}^2}\right).
\end{equation}
One can show that only this last term in expansion of the exponent
will survive in the coincidence limit $\theta^{+}=\theta^{+ '}$
which should be taken in (\ref{proper}), since
$(D^-)^4(D^+)^4\delta^8(\theta-\theta')|_{\theta=\theta'}=1.$ All
other terms with less then four $(D^-)$  are killed at the
coincident limit. As a result, we obtain, as the coefficient the
maximally admissible number of the quantities $A^+, \bar{A}^+$
with non-zero Grassmann parity. Then all the other dependence on
$A^+, \bar{A}^+$ in operator exponents (\ref{kern}) must be
omitted and we get the expression for the effective action
\begin{equation}\label{final}
\Gamma^{(1)}_v [V^{++}, q^+] = \frac{i}{2}n(\Upsilon)\int
d\zeta^{(-4)} e^{-s{\cal M}^2_v} K_{Sch}(s) (A^+)^2 (\bar{A}^+)^2
\times\end{equation}$$ \times \mbox{tr}\left(\frac{\cosh(s{\cal
N})-1}{{\cal N}^2}\right)\mbox{tr}\left(\frac{\cosh(s\bar{\cal
N})-1}{\bar{\cal N}^2}\right)~,
$$
where $K_{Sch}(s)$ is the superfield Schwinger-type kernel
\cite{schwinger}, \cite{bkt}. The latter is defined as follows
$K_{Sch}(x,x',s)=e^{-\frac{s}{2}{\cal D}^{\dot\alpha\alpha}{\cal
D}_{\dot\alpha\alpha}}\{{\bf 1}\delta^4(x-x')\}$. Now a computation
of this heat kernel and its functional trace is standard (see e.g.
\cite{other}, \cite{km} for details). We write down only the final
result
$$
K_{Sch}(s)=\frac{i}{(4\pi s)^2} \frac{s^2({\cal N}^2-\bar{\cal
N}^2)}{\cosh(s{\cal N})-\cosh(s\bar{\cal N})}~.
$$
Here ${\cal N}$ is given by ${\cal N}=\sqrt{-\frac{1}{2}D^4{\cal
W}^2}$. It can be expressed in terms of the two invariants of the
Abelian vector field ${\cal F}=\frac{1}{4}F^{mn}F_{mn}$ and ${\cal
G}=\frac{1}{4}{}^\star F^{mn}F_{mn}$ as ${\cal N}=\sqrt{2({\cal
F}+i{\cal G})}$.

Relation (\ref{final}) is a final result for the hypermultiplet
dependent low-energy one-loop effective action of the
Heisenberg-Euler type. We remind that the whole background
hypermultiplet is concentrated in ${\cal M}_v^2$. The explicit form
of it is:
\begin{equation}\label{manifest}
\Gamma^{(1)}[V^{++}, q^+]=\frac{1}{(4\pi)^2}n(\Upsilon)\int
d\zeta^{(-4)}du\int_0^\infty \frac{ds}{s^3}e^{-s(\alpha^2(H){\cal
W}\bar{\cal W}+r(\Upsilon)q^{+a}q^-_a)} \times
\end{equation}
$$
\times \frac{\alpha^4(H)}{16}(D^+{\cal W})^2(\bar{D}^+\bar{\cal
W})^2\frac{s^2({\cal N}^2-\bar{\cal N}^2)}{\cosh(s{\cal
N})-\cosh(s\bar{\cal N})}\cdot\frac{\cosh(s{\cal N})-1}{{\cal
N}^2}\cdot\frac{\cosh(s\bar{\cal N})-1}{\bar{\cal N}^2}~.
$$
It is easily to see that the integrand in (\ref{manifest}) can be
expanded in power series in the quantities $s^2{\cal N}^2$,
$s^2\bar{\cal N}^2$. After change of proper time $s$ to $s' {\cal
W}\bar{\cal W}$ we get the expansion in power of $s^{' 2}\frac{{\cal
N}^2}{({\cal W}\bar{\cal W})^2}$ and their conjugate. Since the
integrand of (\ref{manifest}) is already $\sim (D^+{\cal
W})^2(\bar{D}^+\bar{\cal W})^2$, we can change in each term of
expansion the quantities ${\cal N}^2$, $\bar{\cal N}^2$ by
superconformal invariants $\Psi^2$ and $\bar\Psi^2$ \cite{bkt}
expressing these quantities from $ \bar\Psi^2=\frac{1}{\bar{\cal
W}^2}D^4\ln {\cal W}=\frac{1}{2\bar{\cal W}^2}\{\frac{{\cal
N}_\alpha^\beta{\cal N}_\beta^\alpha}{{\cal W}^2}+ {\cal O}(D^+{\cal
W})\} $ and its conjugate. After that, one can show that each term
of the expansion can be rewritten as an integral over the full
${\cal N}=2$ superspace.

It is interesting and instructive to evaluate the leading part of
the effective action (\ref{manifest}). Analysis of (\ref{manifest})
(see the details in \cite{bp}) yields
$$
\Gamma^{(1)}_{\rm lead}=\frac{1}{(4\pi)^2}n(\Upsilon)\int
d\zeta^{(-4)}du \frac{1}{16}\frac{D^+{\cal W}D^+{\cal W}}{{\cal
W}^2}\frac{\bar{D}^+\bar{\cal W}\bar{D}^+\bar{\cal W}}{\bar{\cal
W}^2}\frac{1}{(1-X)^2},
$$
where
\begin{equation}\label{X}
X=\frac{-q^{+a} q^-_{a}}{{\cal W}\bar{\cal
W}}\frac{r(\Upsilon)}{\alpha^2({H})}\,.
\end{equation}
As the next step, we rewrite this expression as:
$$
\frac{1}{(4\pi)^2}\int d\zeta^{(-4)}du\frac{1}{16}\{D^{+2}\ln{\cal
W}\bar{D}^{+2}\ln\bar{\cal W} +\sum_{k=1}^\infty
\frac{1}{k^2(k+1)}D^{+2}\frac{1}{{\cal
W}^k}\bar{D}^{+2}\frac{1}{\bar{\cal
W}^k}(-\frac{r(\Upsilon)q^{+a}q^-_a}{\alpha^2(H)})^k\}
$$
\begin{equation}\label{dilog}
 =\frac{1}{(4\pi)^2}\int d^{12}z du\{\ln{\cal W}\ln\bar{\cal
W}+\sum_{k=1}^\infty \frac{1}{k^2(k+1)}X^k\}.
\end{equation}
That exactly coincides, up to group factor $\Upsilon$ with the
earlier results \cite{bi}, \cite{hyper}, \cite{bp}:
\begin{equation}\label{lead}
\Gamma^{(1)}_{\rm lead}=\frac{1}{(4\pi)^2}n(\Upsilon)\int du
\,d^{12}z\, \left(\ln {\cal W}\ln \bar{\cal W}+
\mbox{Li}_2(X)+\ln(1-X)-\frac{1}{X}\ln(1-X)\right).
\end{equation}
Here $\mbox{Li}_{2}(X)$ is the Euler's dilogarithm function.
Next-to-leading corrections to (\ref{lead}) can also be calculated
\cite{bp}. The remarkable feature of the low-energy effective action
(\ref{lead}) is the appearance of the factor
$r(\Upsilon)/\alpha({H})$ in argument $X$. This factor is
conditioned by the vacuum structure of the model under consideration
and depends on the specific features of the symmetry breaking. The
form of the hypermultiplet dependent effective action analogous to
(\ref{lead}) has been found originally in Ref. \cite{bi} for ${\cal
N}=4$ SYM theory and studied in Refs. \cite{hyper}, \cite{bp} by
different methods. In ${\cal N}=4$ SYM theory, the ${\cal N}=2$
vector multiplet and hypermultiplet both belong to the adjoint
representation of the gauge group and the above factor in $X$ is
equal to $1$. As a result, we conclude that the hypermultiplet
dependent low-energy effective action has the universal form
(\ref{lead}) for all ${\cal N}=2$ superconformal models, the
difference of one ${\cal N}=2$ superconformal model from the others
is conditioned only by factor $r(\Upsilon)/\alpha({H})$ in the
quantity $X$ (\ref{X}). The same conclusion concerns also the
general expression (\ref{final}). Of course, the different models
evidently contain the different common factors $n(\Upsilon)$ in
front of integrals (\ref{final}), (\ref{lead}).

Now we discuss some terms in the component Lagrangian corresponding
to the effective action (\ref{lead}). Component structure of the
effective action (\ref{lead}) has been studied \cite{bi} in the
context of  ${\cal N}=4$ SYM theory in bosonic sector for completely
constant background fields $F_{mn}, \phi, \bar\phi, f^i, \bar{f}_i$.
However, it was pointed out above that the superfield effective
action (\ref{lead}) allows us to find the terms in the effective
action up to fourth order in space-time derivatives of component
fields. Now our aim is to find such terms in the hypermultiplet
scalar component sector. To do that we omit all components of the
background superfields besides the scalars $\phi, {\bar{\phi}}$ in
the ${\cal N}=2$ vector multiplet and scalars $f, \bar{f}$ in the
hypermultiplet and integrate over
$d^4\theta^{+}d^4\theta^{-}=(D^-)^4(D^+)^4$. Then we act these
derivatives on the series under the integral in (\ref{dilog}). To
get the leading space-time derivatives of the hypermultiplet scalar
components we should put exactly two spinor derivatives on each
hypermultipelt superfiled. It yields, after some transformations, to
the following term with four space-time derivatives on $q^{\pm}$ in
component expansion of effective action (\ref{dilog}):
\begin{eqnarray}
\Gamma^{(1)}_{\rm lead}&=&\int
d^4xdu\frac{n(\Upsilon)}{(4\pi)^2}\sum_{k=2}^\infty\frac{1}{16}\frac{k-1}{k(k+1)}
\frac{X^{k-2}}{({\cal W}\bar{\cal
W})^2}\{-\bar{D}^{+\dot\alpha}D^{+\alpha}q^-_b
\bar{D}^+_{\dot\alpha}D^-_\beta
q^{+(b}\bar{D}^{-\dot\beta}D^{-\beta}q^{+a)}\bar{D}^-_{\dot\beta}D^+_\alpha
q^-_a
  \nonumber\\
 &+&\frac{1}{2}\bar{D}^{+\dot\alpha}D^{+\alpha}q^-_b
\bar{D}^{-\dot\beta}D^{-\beta}
q^{+b}\bar{D}^-_{\dot\beta}D^-_\beta
q^{+a}\bar{D}^+_{\dot\alpha}D^+_\alpha q^-_a
 \nonumber\\
&+&\frac{1}{2}\bar{D}^{-\dot\beta}D^{+\alpha}q^-_b
\bar{D}^{+\dot\alpha}D^{-\beta}
q^{+b}\bar{D}^+_{\dot\alpha}D^-_\beta
q^{+a}\bar{D}^-_{\dot\beta}D^+_\alpha q^-_a\}|_{\theta=0}\,.
 \nonumber
 \end{eqnarray}
The straightforward calculation of the components\footnote{Here we
have used the relations
$\bar{D}^+D^+q^-=\bar{D}^+D^+D^{--}q^+=-\bar{D}^+D^-q^+=-2i\sigma^\mu\partial_\mu
q^+$, as well as $\bar{D}^-D^-q^+=2i\sigma^\mu\partial_\mu q^-$,
$\bar{D}^+D^-q^+=2i\sigma^\mu\partial_\mu q^+$,
$\bar{D}^-D^+q^-=-2i\sigma^\mu\partial_\mu q^-$. In addition we
have used $\int du
u^+_iu^+_ju^-_ku^-_l=\frac{1}{6}(\epsilon_{ik}\epsilon_{jl}+\epsilon_{il}\epsilon_{jk})$.}
in this expression shows that among the many terms with four
derivatives there is an interesting term of the special type
\begin{equation}\label{chern}
\Gamma^{(1)}_{\rm lead} = \frac{-1}{8\pi^2}
n(\Upsilon)\left(\frac{r(\Upsilon)}{\alpha({
H})}\right)^2[\frac{X_0-2}{X_0^3}\ln(1-X_0)-\frac{2}{X_0^2}]\times\end{equation}$$\times
\int du d^4x
\frac{1}{(\phi\bar\phi)^2}i\varepsilon^{\mu\nu\lambda\rho}\partial_\mu
\tilde{q}^+
\partial_\nu q^+ \partial_\lambda \tilde{q}^-
\partial_\rho q^- ~.$$
As the first term in expansion over variable
$X_0=\frac{r(\Upsilon)\bar{f}^if_i}{\alpha^2\bar\phi\phi}$ we have
\begin{equation}\Gamma^{(1)}_{\rm lead}=-\frac{1}{48\pi^2}
n(\Upsilon)\left(\frac{r(\Upsilon)}{\alpha({H})}\right)^2
\int
d^4x
\frac{1}{(\phi\bar\phi)^2}i\varepsilon^{\mu\nu\lambda\rho}(\partial_\mu
\bar{f}^i\partial_\nu f_i \partial_\lambda \bar{f}^j \partial_\rho
f_j -\partial_\mu \bar{f}^i\partial_\nu \bar{f}_i \partial_\lambda
{f}^j \partial_\rho f_j)\,.
\end{equation}
Here we have omitted all the terms containing the expressions of the
type $\partial^{\mu}f\partial_{\mu}f.$ The expression (\ref{chern})
has a form of the Chern-Simons-like action for the multicomponent
complex scalar filed. The terms of such form in the effective action
were discussed in Refs. \cite{tseyt}, \cite{arg} in context of
${\cal N}=4,2$ SYM models and in Refs. \cite{intr} for $d=6, {\cal
N}=(2,0)$ superconformal models respectively. Here the expression
(\ref{chern}) is obtained as a result of straightforward calculation
in the supersymmetric quantum field theory.

As the examples we list the values of $\alpha({H}), r(\Upsilon)$ and
$n(\Upsilon)$ for models considered in \cite{9}.

$({\bf i})$ ${\cal N}=4$ SYM theory with gauge groups $\mathop{\rm
SU}(N)$, $\mathop{\rm Sp}(2N)$ and $\mathop{\rm SO}(N)$. Here the
hypermultiplet sector is composed of a single hypermultiplet in the
adjoint representation of the gauge group. The background was chosen
such that the gauge groups are broken down as follows $\mathop{\rm
SU}(N)\rightarrow \mathop{\rm SU}(N-1)\times \mathop{\rm U}(1)$,
$\mathop{\rm Sp}(2N)\rightarrow \mathop{\rm Sp}(2N-2)\times
\mathop{\rm U}(1)$, $\mathop{\rm SO}(N)\rightarrow \mathop{\rm
SO}(N-2)\times \mathop{\rm U}(1)$. All background fields aligned
along element $H=\mathop{\rm U}(1)$ of the Cartan subalgebra (with
$\Upsilon=H$). The mass matrix becomes
$$
({\cal M}^2_v)_{IJ}=( {\cal W}\bar{\cal W}+ {\bf q}^{+a}{\bf
q}^-_a)(\alpha(H))^2 \delta_{I,J}
$$
and traces in Eq.(\ref{gamma}) produce the coefficient $n(\Upsilon)$
which is equal to the number of roots with $\alpha(H)\neq 0$, i.e.
to the number of broken generators
$$
n(\Upsilon)=\cases{ 2(N-1)&\hbox {for} $\mathop{\rm SU}(N)$\,,\cr
4N-2&\hbox{for} $\mathop{\rm Sp}(2N)~{\rm and}~\mathop{\rm
SO}(2N+1)$\,,\cr 4N-1&\hbox{ for} $\mathop{\rm SO}(2N)$\,.\cr}
$$
The form of the mass matrix shows that in this case $r(\Upsilon) =
\alpha(H)$ .

$({\bf ii})$ The model introduced in \cite{Ahar}. The gauge group is
$\mathop{\rm USp}(2N)=\mathop{\rm Sp}(2N, \mathbb{C})\bigcap
\mathop{\rm U}(2N)$. The model contains four hypermultiplets $q^+_F$
in the fundamental and one hypermultiplet $q^+_A$ in the
antisymmetric traceless representation $\mathop{\rm USp}(2N)$. The
background fields ${\cal W}$, $q^+_F$, $q^+_A$ are chosen to solve
Eqs. (\ref{vacua}) with the unbroken maximal gauge subgroup
$\mathop{\rm USp}(2N-2) \times \mathop{\rm U}(1)$:
$$
{\cal W}=\frac{{\bf \cal
W}}{\sqrt{2}}\,\mbox{diag}(1,\underbrace{0,...,0}_{N-1},
-1,\underbrace{0,...,0}_{N-1}), \quad q^+_F=0\,,
$$
$$
(q^+_A)_\alpha^{\;\;\beta}=\frac{\bf
q^+}{\sqrt{2N(N-1)}}\,\mbox{diag}(N-1,\underbrace{-1,...,-1}_{N-1},
N-1,\underbrace{-1,...,-1}_{N-1})\,.
$$
The mass matrix $({\cal M}^2_v)_{IJ}$ has been calculated in
\cite{9} and it has $n(\Upsilon)=4(N-1)$ eigenvectors with the
eigenvalue ${\cal M}^2_v=\bar{\bf \cal W}{\bf \cal
W}+\frac{N}{N-1}\bar{\bf q}^j{\bf q}_j$.

$({\bf iii})$ The ${\cal N}=2$ superconformal model which is the
simplest quiver gauge theory \cite{kach}, \cite{doug}. Gauge group
is $\mathop{\rm SU}(N)_L\times \mathop{\rm SU}(N)_R$. The model
contains two hypermultiplets $q^+$, $\tilde{q}^+$ in the
bifundamental representations $({\bf N},\bar{\bf N})$ and $(\bar{\bf
N}, {\bf N})$ of the gauge group. In \cite{9} a solutions of
(\ref{vacua}) with non-vanishing hypermultiplet components that
specifies the flat directions in massless ${\cal N}=2$ SYM theories
has been constructed. The moduli space of vacua for this model
includes the following field configuration
$$
{\cal W}_L={\cal W}_R=\frac{ {\cal
W}}{N\sqrt{2(N-1)}}\,\mbox{diag}(N-1,
\underbrace{-1...,-1}_{N-1})\,,
$$
$$
q^+=\tilde{q}^+=\frac{{\bf
q}^+}{\sqrt{2}}\,\mbox{diag}(1,0,...,0)\,,
$$
which preserves an unbroken gauge group $\mathop{\rm SU}(N-1)\times
\mathop{\rm SU}(N-1)$ together with the diagonal $\mathop{\rm U}(1)$
subgroup in $\mathop{\rm SU}(N)_L\times \mathop{\rm SU}(N)_R$
associated with the chosen ${\cal W}$. In such a background the mass
matrix  has eigenvalue ${\cal M}^2_v=\frac{1}{N-1}\bar{\cal W}{\cal
W} +\frac{1}{N}{\bf q}^{+a}{\bf q}^-_a$ and the corresponding
$n(\Upsilon)=4(N-1)$.

\section{Hypermultiplet dependent contribution \\to
the effective action beyond the on-shell condition}

In the above consideration, as well as in the papers on the
hypermultiplet dependent effective action \cite{bi}, \cite{hyper}
\cite{bp}, a crucial point was the condition that the hypermultiplet
$q^+$ satisfies the one-shell conditions (\ref{onsh}) and the
constraint $q^+=D^{++}q^-$. As it has been pointed out in \cite{bi}
these conditions are sufficient to get all the leading low-energy
contributions to the effective action. Here we relax the on-shell
conditions and study some of possible subleading contributions with
the minimal number of space-time derivatives in the component
effective action.

\begin{figure} [ht]
\begin{center}
\includegraphics{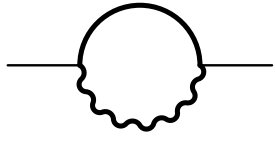}
\caption{One-loop supergraph} \label{f1}
\end{center}
\end{figure}

We consider a supergraph given in Fig.\ref{f1} with two external
hypermultiplet legs and with all propagators depending on the
background ${\cal N}=2$ vector multiplet. Here the wavy line stands
for the ${\cal N}=2$ gauge superfield propagator and the solid
external and internal lines stand for the background hypermultiplet
superfields and quantum hypermultiplet propagator respectively. For
simplicity we suppose that the background field is Abelian and omit
all group factors. The corresponding contribution to effective
action looks like
\begin{equation}\label{gamma_2}
i\Gamma_2 =\int d \zeta_1^{(-4)} d \zeta_2^{(-4)} du_1
du_2\left(\frac{({\cal D}^+_1)^4 ({\cal D}^+_2)^4}{(u^+_1
u^+_2)^3}\frac{1}{{\stackrel{\frown}{\Box}}_1}\delta^{12}(1|2)\right)
\times
\end{equation}
$$
\times \left(\frac{({\cal D}^+_2)^4 ({\cal
D}^+_1)^4}{{\stackrel{\frown}{\Box}}_2{\stackrel{\frown}{\Box}}_1}
\delta^{12}(2|1)({\cal D}_1^{--})^2
\delta^{(-2,2)}(u_2,u_1)\right)
\tilde{q}^{+}(z_1,u_1)q^+(z_2,u_2).
$$
As usually, we extract the factor $(D^+)^4$ from the vector
multiplet propagator for reconstructing the full ${\cal N}=2$
measure. Then we shrink a loop into a point by transferring the
$\stackrel{\frown}{\Box}$ and $({\cal D}^+)^4$ from first
$\delta$-function to another one and kill one integration. At this
procedure the operator $\stackrel{\frown}{\Box}$ does not act on
$q^+$ because we are interesting in the minimal number of space-time
derivatives in the component form of the effective action. As a
result, one obtains
$$
i\Gamma_2 =\left.\int \frac{d \zeta_1^{(-4)}  du_1 du_2}{(u^+_1
u^+_2)^3} \frac{({\cal D}^+_1)^4({\cal D}^+_2)^4 ({\cal
D}^+_1)^4}{{\stackrel{\frown}{\Box}}_2{\stackrel{\frown}{\Box}}_1^2}
\delta^{12}(z-z')\right|\times
$$
$$
\times \left( ({\cal D}_1^{--})^2 \delta^{(-2,2)}(u_2,u_1)\right)
\tilde{q}^+(z_1,u_1)q^+(z_1,u_2)~.
$$
Further we use twice the relation (\ref{polin}) \cite{skim} allowing
us to express the $({\cal D}^{+}_1)^4({\cal D}^+_2)^4$ as a
polynomial in powers of $(u^+_1u^+_2)$.  Then after multiplying the
$({\cal D}^+_1)^4({\cal D}^+_2)^4 ({\cal D}^+_1)^4$ with the
distribution $1/(u^+_1u^+_2)^3$ we obtain a polynomial in
$(u^+_1u^+_2)$ containing the powers of this quantity from 5-th to
1-st. The first order is just a contribution of the type which we
considered in the previous section, because one derivation
$(D^{--})^2$ is used for transformation $(u^+_1u^+_2)$ into
$(u^+_1u^-_2)|_{u_1=u_2} =1$ in the coincident limit. Another
$D^{--}$ transforms $q^+$ into $q^-$. All that has been already done
in Section 4. Therefore, keeping only the first order in
$(u^+_1u^+_2)$ we get a contribution including the combination
$q^{+}q^{-}$ without derivatives. As we pointed out in Section 4, to
obtain such a contribution it is sufficient to consider the
hypermultiplet satisfying on-shell condition (\ref{onsh}).

Here we consider the new contribution to the effective action
containing term $(u^+_1u^+_2)^2$ in the above polynomial:
\begin{equation}
\frac{({\cal D}^+_1)^4({\cal D}^+_2)^4 ({\cal D}^+_1)^4}{(u^+_1
u^+_2)^3}=\end{equation}$$...+ (u^+_1
u^+_2)^2(u^-_1u^+_2)(u^-_2u^+_1)({\cal
D}^+_1)^4\left(\frac{i}{2}{\stackrel{\frown}{\Box}}_1
\Delta^{--}_2 (u^+_2u^-_1)-\frac{i}{2} \Delta^{--}_1
{\stackrel{\frown}{\Box}}_2 (u^+_1u^-_2)\right)+...
$$
The ellipsis means the terms with the powers of $(u^+_1u^+_2)$
other then 2. One can show that in the coincident limit they
disappear. Now transferring $(D^{--})^2$ on $(u^+_1u^+_2)^2$ we
obtain the expression:
\begin{equation}
i\Gamma_2 = i \int d\zeta^{(-4)}du ({\cal D}^+)^4
\frac{1}{{\stackrel{\frown}{\Box}}^3}(\underbrace{{\stackrel{\frown}{\Box}}\Delta^{--}}_{\Gamma_2(1)}-
\underbrace{\Delta^{--}{\stackrel{\frown}{\Box}}}_{\Gamma_2(2)})\delta^{12}(z-z')|_{z=z'}\tilde{q}^+(z,u)q^+(z,u)\,,
\end{equation}
where $\Delta^{--}$ is defined in (\ref{Delt}).

Let us consider each of the two underlined contributions
separately. We use the representation
\begin{equation}\label{repr}
\frac{1}{{\stackrel{\frown}{\Box}}^2}\Delta^{--}\delta^{12}(z-z')|=\int
ds\, s e^{s{\stackrel{\frown}{\Box}}}\Delta^{--}
\delta^{12}(z-z')|,
\end{equation}
where $|$ means the coincident limit $z=z'$. Then we  can apply a
derivative expansion of the heat kernel. The goal is to collect the
maximum possible number of factors of ${\cal D}^+, {\cal D}^-$
acting on $(\theta^+ -\theta^{'+})^4(\theta^- -\theta^{'-})^4$ and
having the minimum order in $s$ in the integral over $s$. Higher
orders in $s$ generate the higher spinor derivatives in the
effective action. We take terms $\frac{1}{2}{\cal W}({\cal D}^-)^2
+c.c.$ from $\Delta^{--}$ and expand the exponential so as to find
$({\cal D}^-)^4$. The Eq. (\ref{repr}) allows us to write the
leading contribution to $\Gamma_2(1)$ as follows
$$ \Gamma_2(1)=-\int d^{12}zdu\int^\infty_0ds \cdot s\int
\frac{d^4p}{(2\pi)^4}e^{-sp^2}e^{s({\cal W}\bar{\cal
W}-\varepsilon)}\frac{s^2}{32} \bar{\cal W}(D^{+\alpha}{\cal
W}D^+_\alpha{\cal W})\times
$$$$\times (D^-)^2(\bar{D}^-)^2\delta^8(\theta-\theta')| \tilde{q}^+ q^+ +
\hbox{ c.c.}
$$
After trivial integration over p and s this contribution has the
form
\begin{equation}\label{gamma(1)}
\Gamma_2(1)=\frac{i}{32\pi^2}\int d^{12}z du \frac{1}{\bar{\cal
W}}\frac{D^+{\cal W}D^+{\cal W}}{{\cal W}^2}
\tilde{q}^+(z,u)q^+(z,u)({\cal D}^-)^4 \delta^{8}(\theta-\theta')|
\end{equation}
$$+\frac{i}{32\pi^2}\int d^{12}z du
\frac{1}{{\cal W}}\frac{\bar{D}^+\bar{\cal W}\bar{D}^+\bar{\cal
W}}{\bar{\cal W}^2} \tilde{q}^+(z,u)q^+(z,u)({\cal D}^-)^4
\delta^{8}(\theta-\theta')|\,.
$$
Now we fulfil the same manipulations with the second underlined
contribution $\Gamma_2(2)$ keeping the same order in $s$ and $D^-,
\bar{D}^-$ as in the expression (\ref{gamma(1)}):
$$
\Gamma_2(2)=-\int d^{12}zdu\tilde{q}^+q^+\int_0^\infty \frac{ds
s^2}{2}\int\frac{d^4p}{(2\pi)^4}e^{-sp^2+isp{\cal
D}+s\stackrel{\frown}{\Box}}\times
$$
$$
\times(ip^{\alpha\dot\alpha}{\cal D}^-_\alpha\bar{\cal
D}^-_{\dot\alpha}+\Delta^{--})(-\frac{1}{2}p^{\alpha\dot\alpha}p_{\alpha\dot\alpha}+ip^{\alpha\dot\alpha}{\cal
D}_{\alpha\dot\alpha}+\stackrel{\frown}{\Box})\delta^8(\theta-\theta')|
$$
$$
=-\int d^{12}zdu\tilde{q}^+q^+\int_0^\infty \frac{ds
s^2}{2}\int\frac{d^4p}{(2\pi)^4}e^{-sp^2 +s{\cal W}\bar{\cal
W}-\varepsilon s}\times$$$$\times\frac{1}{2}\bar{\cal
W}(\bar{D}^-)^2\frac{1}{4}(D^+{\cal W})(D^+{\cal
W})\{-\frac{s^2}{4}p^{\alpha\dot\alpha}p_{\alpha\dot\alpha}+s\}\frac{1}{2}(D^-)^2\delta^8(\theta-\theta')|
+({\cal W}\leftrightarrow \bar{\cal W})~.
$$
Integration over momenta in this
expression\footnote{$\int\frac{d^4p}{(2\pi)^4}e^{-sp^2}=\frac{i}{(4\pi
s)^2}$,
$\int\frac{d^4p}{(2\pi)^4}p_{\alpha\dot\alpha}p^{\beta\dot\beta}e^{-sp^2}=\frac{i}{(4\pi
s)^2}\frac{1}{s}\delta^{\beta\dot\beta}_{\alpha\dot\alpha}$} gives
$\frac{i}{(4\pi s)^2}\{-\frac{s^2}{4}\frac{4}{s}+s\}=0~.$ After that
we see that the leading term of the form (\ref{gamma(1)}) is absent
in $\Gamma_2(2)$. Then it is not difficult to show that the
contribution (\ref{gamma(1)}) is rewritten as follows [we use $\int
d^2\bar\theta^-=\bar{D}^{+2}$]
$$
-\frac{i}{32\pi^2}\int d^4x d^4\theta^+ d^2\theta^- du (\bar{D}^+)^2
(D^+)^2( \ln{\cal W}) \frac{1}{\bar{\cal
W}}\tilde{q}^+(z,u)q^+(z,u)({\cal D}^-)^4
\delta^{8}(\theta-\theta')|+\hbox{ c.c.}
$$
The non-zero result arises when all $D^{+}$ - factors act only on
the spinor delta-function. Thus, the contribution under
consideration is written as an integral over the measure
$d^4xdud^4{\theta}^+d^2{\theta}^-$ which looks like
 "$3/4$ - part" of the full ${\cal N}=2$ harmonic superspace measure
$d^4xdud^4{\theta}^+d^4{\theta}^-$.

Therefore, the hypermultiplet dependent effective action contains
the term
\begin{eqnarray}\label{3/4}
\Gamma_2=&-&\frac{i}{32\pi^2}\int d^4xdud^4\theta^+ d^2\theta^-
\frac{1}{\bar{\cal W}} \ln ({\cal W})\tilde{q}^+q^+ |_{\bar\theta^-=0}\\
&-&\frac{i}{32\pi^2}\int d^4xdud^4\,\theta^+ d^2\bar\theta^-
\frac{1}{{\cal W}} \ln (\bar{\cal
W})\tilde{q}^+q^+|_{\theta^-=0}~.\nonumber
\end{eqnarray}
Presence of such a term in the effective action for ${\cal N}=2$
supersymmetric models in subleading order was proposed in
\cite{arg}. Here we have shown how this term can be derived in the
supersymmetric quantum field theory.

It is interesting and instructive to find a component form of such a
non-standard superfield action (\ref{3/4}). Here we consider only a
purely bosonic sector of (\ref{3/4}). After integration over
anticommuting variables, which can be equivalently replaced by
supercovariant derivatives evaluated at $\theta=0$, one gets:
$$
\Gamma_2=\frac{i}{4\pi^2}\int d^4xdu\frac{1}{{\cal W}\bar{\cal
W}}D^+_\beta\bar{D}^-_{\dot\alpha}\tilde{q}^+\bar{D}^{-\dot\alpha}D^-_\alpha
q^+D^{+\beta}D^{-\alpha}{\cal
W}|_{\theta,\bar\theta=0}$$$$+\frac{i}{4\pi^2}\int
d^4xdu\frac{1}{{\cal W}\bar{\cal
W}}\bar{D}^+_{\dot\beta}{D}^{-\alpha}\tilde{q}^+{D}^-_\alpha\bar{D}^-_{\dot\alpha}
q^+\bar{D}^{+\dot\beta}\bar{D}^{-\dot\alpha}\bar{\cal
W}|_{\theta,\bar\theta=0}.
$$
Since
$D^-_\alpha\bar{D}^-_{\dot\alpha}q^+=-2i\partial_{\alpha\dot\alpha}f^-
$, $\bar{D}^+_{\dot\beta}D^-_\beta
\tilde{q}^+=2i\partial_{\beta\dot\beta}\tilde{f}^+$ we have
\begin{equation}\Gamma_2=\frac{i}{\pi^2}\int d^4x du
\frac{1}{\phi\bar\phi}\partial_{\beta\dot\alpha}\tilde{f}^+
\partial^{\dot\alpha}_\alpha f^- F^{\beta\alpha} + \frac{i}{\pi^2}\int d^4x du
\frac{1}{\phi\bar\phi}\partial_{\dot\beta}^{\alpha}\tilde{f}^+
\partial_{\alpha\dot\alpha} f^-
\bar{F}^{\dot\beta\dot\alpha},
\end{equation}
where $F^{\alpha\beta}, {\bar{F}}^{\dot{\alpha}\dot{\beta}}$ are the
spinor components of Abelian strength $F_{ab}$. Then one converts
the spinor indices into vector ones. As a result, we obtain a
Chern-Simons-like contribution to the effective action containing
three space-time derivatives
\begin{equation}\label{last}
\Gamma_2= -\frac{1}{2\pi^2}\int d^4x
\frac{1}{\phi\bar\phi}\varepsilon^{mnab}\partial_m \bar{f}^i
\partial_n f_i F_{ab}~.
\end{equation}
This expression is the simplest contribution to the hypermultiplet
dependent effective action beyond the on-shell conditions
(\ref{onsh}) for the background hypermultiplet. Of course, there
exist other, more complicated contributions including the
hypermultiplet derivatives, they also can be calculated by the same
method which led to (\ref{3/4}). Here we only demonstrated a
procedure which allows us to derive the contributions to the
effective action in the form of integral over $3/4$ - part of the
full ${\cal N}=2$ harmonic superspace.

\section{Summary}
We have studied the one-loop low-energy effective action in ${\cal
N}=2$ superconformal models. The models are formulated in harmonic
superspace and their filed content correspond to the finiteness
condition (\ref{fin}). Effective action depends on the background
Abelian ${\cal N}=2$ vector multiplet superfield and background
hypermultiplet superfields satisfying the special restrictions
(\ref{vacua}), (\ref{vac}) which define the vacuum structure of the
models. The effective action is calculated on the base of the ${\cal
N}=2$ background field method for the background hypermultiplet
on-shell (\ref{onsh}) and beyond the on-shell conditions.

We have shown that the hypermultiplet dependent one-loop effective
action for the theory under consideration is associated with a
special superfield operator (\ref{operator}), acting only in the
sector of quantum vector multiplet superfields. The coefficients of
this operator contain the background superfields and depend on
details of gauge symmetry breaking. We prove that for evaluating the
one-loop effective action it is sufficient to consider the simple
case when the  operator has the universal form (\ref{univer}).

The hypermultiplet dependent one-loop low-energy effective action is
calculated in the form of an integral over the proper time. It was
proved that to find the low-energy contributions to the effective
action it is sufficient to consider on-shell vector multiplet and
hypermultiplet. Final result for such a case is given by the
relation (\ref{final}) which is the ${\cal N}=2$ superfield analog
of the Heisenberg-Euler effective action. The leading part of the
low-energy effective action (\ref{lead}) has a general
form\footnote{The form of leading low-energy effective
action(\ref{lead}) has been first found in \cite{bi} for ${\cal
N}=4$ SYM theory.} and depends on the quantity $X=\frac{-{\bf
q}^{+a} {\bf q}^-_{a}}{{\cal W}\bar{\cal
W}}\frac{r(\Upsilon)}{\alpha({H})}$ (\ref{X}) containing the details
of the vacuum structure of the model. Using the superfield effective
action (\ref{lead}) we calculated the lowest space-time dependent
terms in the sector of scalar hypermultiplet components. These terms
contain four space-time derivatives of scalar fields and have a
Chern-Simons-like form (\ref{chern}).

We studied possible contributions to the effective action which can
be generated if we go beyond on-shell conditions (\ref{onsh}) for
the background hypermultiplet. The harmonic supergraph with two
external hypermultiplet legs and with background vector multiplet
dependent propagators has been computed and its leading low energy
contribution has been found. We proved that the final result has a
very interesting superfield structure and is written as an integral
over $3/4$ of the full ${\cal N}=2$ harmonic superspace (\ref{3/4}).
The presence of such terms in the effective action of ${\cal N}=2$
supersymmetric theories was recently proposed in \cite{arg}. We
computed the component structure of the effective action (\ref{3/4})
in the bosonic sector keeping the scalar components of the
background hypermultiplet and vector component of the background
${\cal N}=2$ gauge multiplet. The result (\ref{last}) has a
Chern-Simons-like form and contains three space-time derivatives of
the component fields.

To conclude, we have analyzed the general structure of the
hypermultiplet dependent one-loop low-energy effective action of
${\cal N}=2$ superconformal models. For an on-shell hypermultiplet
we found the universal expression for the effective active action.
For hypermultiplet beyond on-shell, we calculated the special
manifestly ${\cal N}=2$ supersymmetric subleading contribution which
is written as an integral over $3/4$ of the full ${\cal N}=2$
harmonic superspace. We believe that such contributions deserves a
special study.

\section*{Acknowledgments}
N.G.P is grateful to  S. Kuzenko and I. McArthur for helpful
discussions and correspondence. The authors are grateful to A.
Banin and S. Kuzenko for reading the manuscript and useful
remarks. The work was supported in part by RFBR grant, project No
06-02-16346, grant for LRSS, project No 4489.2006.2 and INTAS
grant, project  No 05-7928. I.L.B is grateful to DFG grant,
project No 436 RUS 113/669/0-3 and joint RFBR-DFG grant, project
No 06-02-04012 for partial support. The work of N.G.P was
supported in part by RFBR grant, project No 05-02-16211. I.L.B is
thankful to FAPEMIG (Minas Gerais, Brazil) for finance support
during his stay in Universidade Federal de Juiz de Fora, MG,
Brazil and Prof. I.L Shapiro for kind hospitality.

\end{document}